\documentclass[10pt]{article} 
\usepackage{graphicx,graphics}
\usepackage{blindtext}
\usepackage{geometry}
\usepackage{epsfig}
\usepackage{multicol}
\usepackage{authblk}
\usepackage{kpfonts}
\usepackage{amsmath}
\usepackage{bm}
\usepackage{arydshln}
\title{Toward Understanding of the Role of Reversibility of Phenotypic Switching in the Evolution of Resistance to Therapy.}
\author[1]{D. Horvath}
\author[2]{B. Brutovsky\vspace{2mm}}
{\it \affil[1]{{\it Center of Interdisciplinary Biosciences, 
Technology and Innovation Park, P.~J.~\v{S}af\'arik University,
Jesenn\'a 5, 04154 Ko\v{s}ice, Slovak Republic \vspace{5mm}}}
\affil[2]{{\it Department of Biophysics, Faculty of Science, Jesenn\'a 5,
P.~J.~\v{S}af\'arik University, 04154 Ko\v{s}ice, Slovak Republic}}}

\date{}
\begin{document}

\maketitle

\begin{abstract}
Reversibility of state transitions is intensively studied topic
in many scientific disciplines over many years. In cell biology,
it plays an important role in epigenetic variation of phenotypes,
known as phenotypic plasticity. More interestingly, the cell state
reversibility is probably crucial in the adaptation of population
phenotypic heterogeneity to environmental fluctuations by evolving
bet-hedging strategy, which might confer to cancer cells resistance
to therapy. In this article, we propose a formalization of the evolution
of highly reversible states in the environments
of periodic variability.
Two interrelated models of heterogeneous cell populations are
proposed and their behavior is studied. The first model captures
selection dynamics of the cell clones for the respective levels
of phenotypic reversibility. 
The second model focuses on the interplay between reversibility 
and drug resistance in the particular case of cancer.  
Overall, our results show that the threshold 
dependencies are emergent features of the investigated 
model with eventual therapeutic relevance. 
Presented examples 
demonstrate importance of taking into account cell to cell 
heterogeneity within a system of clones with different reversibility 
quantified by appropriately chosen genetic and epigenetic entropy measures.
\end{abstract}

\numberwithin{equation}{section}
\numberwithin{figure}{section}
\numberwithin{table}{section}
\numberwithin{figure}{section}

\begin{multicols}{2}
\section{Introduction}

Human diseases are typically caused by invading pathogenic
microorganisms, such as viruses, bacteria, fungi, prions, etc. 
Despite immunity system is able to cope with
the majority of pathogens, those escaping from innate immunity
surveillance must be treated by therapy.
The main obstacle to efficient therapy is the variability
of the microorganisms within the population, such as strain or clone,
conferring them resistance to therapy.
Population variability flows from the evolutionary essence
of population dynamics of microorganisms which equips evolving
populations of pathogens with powerful adaptive capability.
For example, evolutionary dynamics of carcinogenesis
\cite{Nowell1976,Merlo2006,Greaves2007} is
considered the main reason why targeted therapy does not work
\cite{Gillies2012}.

As it is known for a long time, intratumor 
heterogeneity is not bound exclusively to the 
differences in DNA sequences of the respective 
cells (genetic heterogeneity), but to the 
epigenetic differences as well \cite{Laird2005}. 
It is broadly accepted that the epigenetic mechanisms for gene transcription regulation
are usually reversible \cite{Willbanks2016}. It has been found 
that the role of epigenetic mechanisms, such as DNA methylation, histone
modifications, chromatin remodeling, and small RNA molecules,
in cancer initiation and progression is causative
\cite{Bjornsson2004,Easwaran2014}. In particular, 
variability in phenotypic characteristics
of isogenic cells, known as phenotypic 
plasticity, is assumed to be an important cause of the therapeutic resilience of advanced 
cancers \cite{Greaves2015_CancerDiscov}. 

Recognizing tumor dynamics as evolutionary process, one can exploit
known universal features of evolution to influence the
evolution of the population in desirable direction in mathematically
more purposeful way. While the role of genetic intratumor heterogeneity in tumor evolution
was accepted long time ago, the role of epigenetic heterogeneity
is much less obvious. To implement evolutionary principles into the therapy design,
the interplay between genetic diversity and epigenetic plasticity \cite{Carja2017}
should be carefully studied at the model level within an evolutionary-based
integrative conceptual framework. Recently, the concept of Waddington's epigenetic landscape was applied
to formalize the relation between the cancer cell genome and epigenetic
mechanisms \cite{Huang2013_CancerMetastasisRev} in mathematically 
instructive way. Therein, each point in the fitness landscape (i.~e.~genome)
provides epigenetic landscape of unique topology. Due to their mathematical
complexity, the epigenetic landscapes contain many areas (attractors)
around the stable cell-states corresponding to the respective stable
phenotypes of a cell. Straightforwardly, the phenotypic switching
corresponds to the transition between two such attractors. 

Below we use the above conceptualization
\cite{Huang2013_CancerMetastasisRev}
as an instructive  backbone for our considerations. As different cell states
in the epigenetic landscape differ in their
fitness-related properties, the cell states composition (or non-genetic
heterogeneity) becomes, from the viewpoint of the clone, evolutionary
important at the clone's respective timescale.
It was observed that in the case of variable selective pressure,
population of organisms evolve mechanisms to tune the phenotypic variability
to the variability of the acting selective pressure \cite{Rando2007}.
In bacteria, the well known risk-diversification strategy evolved in the
populations when facing uncertain future and/or environment
\cite{Crean2009,Forbes2009,Beaumont2009} is the bet-hedging
strategy \cite{DeJong2011,Donaldson2008,Carja2017}.
Based on formal similarity of evolving cancer cells population with
bacteria, viruses or yeast, it has been recently proposed that the structure of intratumor
heterogeneity is evolutionary trait as well, evolving to maximize clonal
fitness at a cancer-relevant timescale in changing (or uncertain)
environment and that its structure corresponds to the bet-hedging
strategy \cite{Brutovsky2013,Nichol2016,Horvath2016,Gatenby2017,Thomas2017}
which has been recently put into therapeutic context \cite{Chisholm2016,Mathis2017}.
To sum up, the genome stays the main protagonist (i.e. selection unit)
in the evolution of cancer cells, nevertheless with 
non-genetic heterogeneity of its eventual clone being the crucial adaptive 
trait at cancer-relevant, instead of proximate timescale.

Phenotypic plasticity confers to cellular tissues important
properties, such as the ability of cancer cells to escape
targeted therapy by switching to an alternative phenotype
\cite{Kemper2014,Rogers2014,Klevebring2014,Pisco2015,Emmons2016,Liu2016_Oncotarget}.
It motivates the effort to stimulate (or prevent) specific phenotype
switching purposefully as a therapeutic strategy \cite{SaezAyala2013},
which requires deep understanding of the phenotype switching causation.

As in classical evolutionary theory are random mutations, which cause phenotypic  variation,
independent of selective pressure, it seems probable that reversibility of phenotype switching can be
underpinned exclusively by epigenetic modifications.
Regarding the therapeutical perspectives, the difference between
genetic and epigenetic changes is fundamental: while the genetic
changes are, in principle, irreversible,
the epigenetic modifications may be reversible at the therapy-relevant
timescales \cite{Rogers2014,Germain2016,Mishra2014}.

To sum up, there are well-founded reasons for studying phenotypic 
switching using quasispecies or population dynamics models,  
which address the specific characteristics of cancer.  
In Section \ref{sec:Quasi} we present the evolutionary 
based framework which assumes environmental dynamics shaping 
the evolution of reversible switching strategies.
In the quasispecies framework (see Subsections 
\ref{sec:Het},\ref{sec:Num}, Appendix \ref{Ap:Long}) 
the focus is placed on the phenotypic plasticity 
and reversibility. Studies of quasispecies models yielded important modelisation 
ideas about how to implement the immunotherapy aspects.  
Most of our efforts in this area have focused on 
the construction of the population-based cancer models where 
evolution of phenotypic reversibility is incorporated 
together with immunotherapy. The resulting 
model of anti-cancer drug resistance are discussed 
in Section \ref{sec:Evo}.  In Subsection \ref{sub:divim} 
we reassess the problem of intratumoral heterogeneity and related entropy 
measures. The corresponding simulation results are provided
in Subsection \ref{sub:SimRes}. 
The results respond to specific questions concerning 
the evolutionary nature of the drug resistance. 
The threshold concept has emerged 
as natural outcome of the simulations. Its importance 
was underlined 
by providing a more detailed 
supplementary analysis in the 
Appendix \ref{Ap:Thresh}.

\section{Quasispecies model of phenotypic 
changes}\label{sec:Quasi}

In this section we introduce the qualitative time-continuous 
evolutionary model based on the system of ordinary 
differential equations. The model is applied to nonequilibrium scenarios,
where constrained populations of irreversible 
phenotypes are evolutionary drawn towards 
an attractor populated by reversible phenotypes.

Recently, mathematical modeling of phenotypic evolution in variable environment 
has received significant attention \cite{Carja2012,Horvath2016,Zheng2017,Frey2017},
which  implicitly addresses the issue of eventual adaptivity of epigenetic
modifications \cite{Herman2014}. In particular, the formulation based on the quasispecies model \cite{eigen89}
has been developed to clarify the role of mutations in the evolutionary process
\cite{Baake1997,Bagnoli98,Malarz98,Bianconi2011}.  
Owing to its versatility, the model has found applications 
in a broad range of scales - from molecular 
and virus scales \cite{Wilke2005} up to the cellular
systems, such as the populations of heterogeneous cancer 
cells \cite{Amor2014,Sardanyes2017}.  
Investigation of heterogeneous malignant 
tissues within the context of variable environments \cite{Maley2017} 
including  local anticancer drug-induced 
environments \cite{Lorz2015} emerges as an interesting research area.

The entropic variable constraints imposed on the heterogeneity 
were studied in \cite{HorvKnel2009}. In \cite{Horvath2010},
the elimination of heterogeneity in the system of replicating
entities was conceived as an inverse problem, with an eventual
potential in therapeutical applications.

Before going into details, we highlight 
three salient features of the model:
(i) periodicity of the (micro)environmental variations;
(ii) substantial genetic diversity underlying the 
switching rates between isogenic phenotypes (phenotypic states);
(iii) competition between the clones adopting 
either reversible or irreversible 
phenotype switching under the 
constraint of constant total cell population size.

Below we study effects of the 
variation of evolutionary rate on the fractions 
$c^{(z)}(k,t),$ $k=0,1, \ldots,$ 
$n_s-1$, of $n_s$ clones (abstract "genotypes"), 
each of them allowed to occupy one of the two phenotypic states, 
indexed by $z\in\{0, 1\}$. We assume 
that the $k$-th genotype responds
to the environmental variation in two alternative ways
depending on its respective phenotypic state, $[k; z=0]$ or $[k; z=1]$.
The time-variability of environment can be indirectly represented by the 
time-variability of the reproduction rates of the respective phenotypic states, 
$r^{(z)}(t)$, $z\in \{0,1\}$, which are supposed to be of harmonic periodic form  
\begin{equation} 
r^{(z)}(t) = r_{B}+ (-1)^z \Delta r  \cos(2 \pi t/ T) 
\label{eq:r01}
\end{equation} 
defined by the period $T$; $\Delta r$ corresponds to the degree of 
diversification of the reproduction fitness and $r_B$ to its basal level.
It is worthy of noting that in \cite{Urtel2017} the oscillatory external 
(temperature) conditions are used to drive the evolution of the class 
of interacting information-carrying molecular replicators with 
the capability of reversible intermodal switches.

Presuming that the replication
dynamics is restricted to the 
concentration plane $c^{(0)}(k,t)\simeq c^{(1)}(k,t)$,  
the parameter $r_B \simeq [ r^{(0)}(t)+r^{(1)}(t)\,]/2$  may be interpreted 
as static effective replication rate. Many natural cycles may imply changes
in the replication rate, which can be, within the context of
cancer research, exemplified by
the cyclic hypoxia-reoxygenation 
exposure within solid 
tumors \cite{Cairns8903}.

To study quasispecies dynamics with purposefully 
specified the reversibility between the phenotypes
$z=0$ and $z=1$ in the respective clones, we define the parameter
$\varphi \in \langle 0, 1 \rangle$, 
via the linear relation
\begin{equation}
\varphi (k,n_s) = \frac{n_s-k-1}{n_s-2}\,, 
\quad k=1,\ldots, n_s-1\,,
\label{varphi}
\end{equation}
which uniformly discretizes {\em commitment} 
of the cell to the respective phenotype.
The labeling of the clones by the index $k$ is introduced to make next
formalization more feasible. Among the possible states 
we highlight the neutral (or "symmetric") 
case $\varphi(n_s/2,n_s) = 1/2$ (equal commitment to either phenotype),
as well as the "boundary" cases $\varphi(1,n_s) = 1$ and 
$\varphi(n_s-1,n_s) = 0$ corresponding to
exclusive commitment to the respective phenotype, which implies
complete irreversibility. In the following, the range of
indices $k$ is completed with $k=0$ which represents the wild clone
(neither mutations nor transitions between phenotypes are applied).
Let $(1-\varphi(k,n_s)) 
c^{(1)}(k,t)$ and $\varphi(k,n_s) c^{(0)}(k,t)$
specify intensity of the transitions from the phenotype $0$ to the phenotype $1$, and vice versa.
The above terms can be combined to describe the switching flow 
 \begin{eqnarray} 
J_{sw}(k,n_s,t)  &=& 
( 1-\varphi(k,n_s) ) c^{(1)}(k,t) 
\label{eq:Jswkt}
\\
&-&  \varphi(k,n_s) c^{(0)}(k,t)\,,
\nonumber
\end{eqnarray}
which, when incorporated into dynamics, 
changes phenotypic fractions corresponding
to $z=0$ and $z=1$ in each of the respective clones $k=1,2,\ldots, n_s-1$. 
In further, the direction of the switching flow $(-1)^z J_{sw}$ 
is controlled by the prefactor $(-1)^z$. 
Assuming non-negativity of the fractions, 
irreversibility of the "boundary" 
species is consistent with 
unidirectionality 
of the "boundary" flows $J_{sw}(1,n_s,t)=-c^{(0)}(1,t)\leq 0$ and 
$J_{sw}(n_s-1,n_s,t)=c^{(1)}(n_s-1,t)\geq 0$. 
This sharply contrasts 
with the central species $k_{\rm central}=n_s/2$, 
where $\varphi(k_{\rm central},n_s) = 1/2$ 
provides
$J_{sw}(n_s/2,t)=$ $(1/2) [ c^{(1)}(n_s/2,t) $ $
-c^{(0)} (n_s/2,t) ]$  
which reflects the fact that both directions of the
flow (i.e. $J_{sw}\geq 0,$ $J_{sw} \leq 0$) 
are allowed for $c^{(z)}(n_s/2,t)$. 
As $J_{sw}(k,n_s,t)$ is constructed without considering 
explicit causal sensoric response \cite{Kobayashi2017}, 
it can be viewed as a population-level
consequence of evolved bet-hedging strategy within the context 
of quasispecies ODE (ordinary differential equations).
Note that the structure of switching 
term on the population level is similar to that used 
to describe the dichotomous switching 
of tumor cells \cite{Fedotov2007}. 
Specific terms of this type related to the transitions between bacterial 
subpopulations were applied as well \cite{Thattai2004}.

We postulate that the initial population is formed 
exclusively by the zero-th clone, which is gradually redistributed
(by mutation mechanisms) among the concurrent clones,
$k=1,2\ldots n_s-1$. If the phenotype switching absents for $k=0$, 
the dynamics of the population, $c^{(z)}(0,t)$, follows
\begin{eqnarray}
\frac{d c^{(z)}(0,t)}{d t} &=& [ 
r^{(z)}(t) - \Phi(t) 
\label{eq:dczk0}
\\
&-& (n_s-1) \mu\, ] c^{(z)}(0,t)
\nonumber
\end{eqnarray}
and the population fraction $k=0$ 
changes exclusively due to irreversible mutations. 
Their impact is modeled by $-(n_s-1) \mu c^{(z)}(0,t)$
terms proportional to the positive coefficient 
$\mu$.  
The additional assumption is, that all 
mutants 
$k=1,2,\ldots, n_s-1$ 
are produced with the same rate $\mu c^{(z)}(0,t)$. 
The competition controls proliferation via the scalar 
$\Phi(t)$ term which is interpreted later.
The Eq.(\ref{eq:dczk0}) 
is considered together with the ODE system
\begin{eqnarray}
\frac{d c^{(z)}(k,t)}{d t} &=& \left[\, r^{(z)}(t)-\Phi(t)\right] c^{(z)}(k,t) 
\label{eq:dczk1}
\\
&+& \mu c^{(z)} (0,t) + m (-1)^z J_{sw}(k,t) \nonumber
\end{eqnarray}
written for $k=1, 2, \ldots, n_s-1$, constructed to describe
combined effect of the  replication, selection and switching. 

The newly introduced parameter $m$ 
controls the intensity of the phenotype 
switching and can be interpreted
as the measure of reversibility. 
The limit conditions under which 
the effect of $m J_{sw} $ 
is significantly weakened 
can be viewed as conditions supporting conservative 
bet-hedging \cite{Gatenby2017}. 
For the sake of simplicity as well
as better understanding 
of the model implications, 
the parameter $m$ is chosen 
to be uniform over the clones.

Regarding the structure, description and interpretation of the 
switching process at the level of ODE, we note that: (i) it follows 
from the Eq.(\ref{eq:dczk1})
that the total rate of the clonal fractions 
$d(c^{(0)}(k,t)+c^{(1)}(k,t) )/dt $ 
loses explicit dependence on $J_{sw}(k,t)$; 
(ii) without mutation and replication terms the 
particular phenotypic 
equilibria can be formed 
when $[J_{sw}(k)]_{equil} \overset{!}{=} 0$, 
which leads to the ratio
$[c^{(1)}(k)/c^{(0)}(k)]_{equil} =\varphi(k,n_s)/[1-\varphi(k,n_s)]$.
The competition among the clones 
is described conveniently using constraint 
of constant overall 
population density 
\begin{equation}
c^{(0)}_{{\mbox{\tiny$\Sigma$}}}(t) + c^{(1)}_{\mbox{\tiny$\Sigma$}}(t)=1 
\label{eq:c0sum}
\end{equation}
with the particular terms 
\begin{equation}
c^{(z)}_{\Sigma}(t)= \sum_{k=0}^{n_s-1} c^{(z)}(k,t)\,, 
 \qquad z=0, 1\,.
\end{equation}
If the sum of the left 
and right hand sides of Eq.(\ref{eq:dczk0}) 
and Eq.(\ref{eq:dczk1}) is carried out,
and the result is compared with 
Eq.(\ref{eq:c0sum}), the scalar correction 
to the reproduction rate may be constructed as
\begin{equation} 
\Phi(t)= 
\sum\limits_{z=0,1} r^{(z)}(t) 
c^{(z)}_{\mbox{\tiny$\Sigma$}}(t)\,.  
\label{eq:Phi}
\end{equation}
In order to reduce the initial asymmetry, we have used in our 
simulations (see Subsection \ref{sec:Num} in below) 
of quasispecies model the initial conditions 
\begin{eqnarray}
c^{(z)}(k,t=0) &=&\frac{\delta_{k,0}}{2}, 
\qquad z\in \{0,1\}\,, 
\label{eq:initc} 
\end{eqnarray} 
where $\delta_{k,0}$ is the Kronecker delta. 
It means, that in the case of the initial demise of the $k= 0$ species,
we expect the repopulation towards the species 
$k  = 1, 2, \ldots, n_s-1$. 
Since the mutations contribute uniformly to the expansion of all remaining clones ($k\neq 0$)
with the same rate $\mu c^{(z)}(0,t)$,
the decisive contribution to the disparity of the 
fractions is expected solely from the switching mechanism
where diversification is guaranteed by the use of $\varphi(k,n_s)$. 

\subsection{Entropy measures \\ of heterogeneity} \label{sec:Het}

The general concept of entropy is used in various scientific fields 
and can therefore be perceived differently according 
to the purposes of the information processing in the respective 
areas of biological research. Shannon entropy is an example 
of the key  systemic measure designed to quantify the information
storage, transfer or system heterogeneity/diversity. 
One key step in this direction is the idea to join a fitness concept 
to information-based characteristics by the 
value of information \cite{Rivoire2011}. 
According to \cite{Palmer2012}, the entropy-based measure 
can be used as a predictive indicator of the evolutionary 
efficiency in avoiding extinction. The measures can quantify 
diversity within the context of phenotypic switching \cite{Kussell2005}
analogously as in presented work. In the field of cancer biology and diagnostics, 
biostatistics, as well as in the studies of cellular composition, 
the entropy can be utilized to monitor and 
characterize dynamics of heterogeneity at many levels. 
Shannon entropy is currently considered the most 
promising robust metrics and the tumor-specific 
imaging biomarker to enable rational prognostic analysis based on the 
data gathered from CT-scans \cite{Dercle2017}.

In our specific context, to characterize selection strength 
and asymptotic behavior at the systemic level, 
we focus on the measures of heterogeneity of evolving population fractions. 
To quantify the dynamic heterogeneity we utilize two 
different measures.
The first of them is Shannon entropy
\begin{equation}
S_{EPIG}(t) = - \sum_{z=0,1}
\sum_{k=0}^{n_s-1} c^{(z)}(k,t) \ln c^{(z)}(k,t)\,,
\label{eq:SEPIG0}
\end{equation}
which reflects 
the epigenetic information. It is obvious, that this form is sensitive to the
arrangement of phenotypic fractions.  
On the contrary, the introduction of measures based on
the total concentration 
\begin{equation}
c(k,t)=c^{(0)}(k,t)+c^{(1)}(k,t)\,
\label{eq:ctotal} 
\end{equation}
or another meaningful algebraic combinations of 
$c^{(0)}$, $c^{(1)}$ 
is a way how to abandon phenotypic 
details (including subtle oscillations).
Therefore, the genetic heterogeneity and variability 
can be better described by  
\begin{equation}
S_G(t) =  - \sum_{k=0}^{n_s-1}  c(k,t)  
\ln (c(k,t))\,.   
\label{eq:SG}
\end{equation}
Both $S_{EPIG}$ and $S_G$ 
measures can be constructed regarding
often used analogy between probability 
and occupancy fraction \cite{Horvath2010}.

\end{multicols}

\begin{figure}[!htb]
\centering
\begin{tabular}{c}
\includegraphics[scale=0.60]{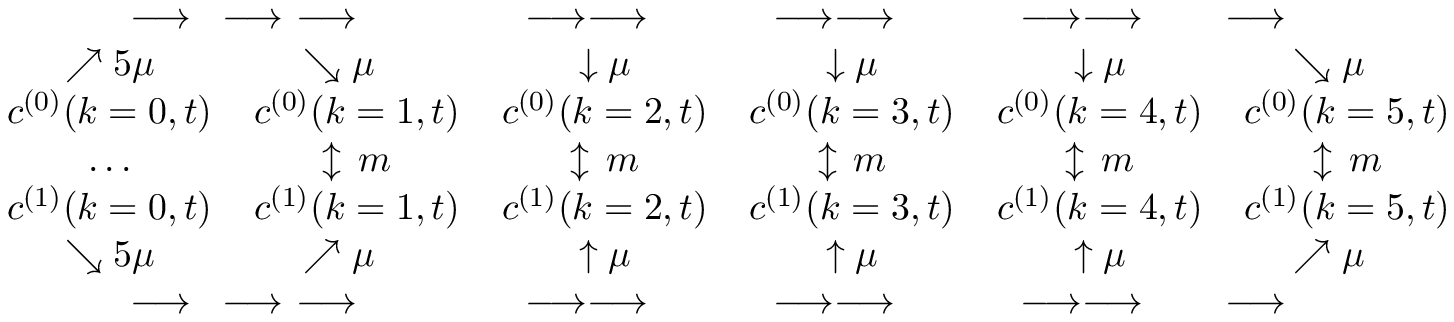}
\\
\vspace*{-3.5mm}
\\
\hspace*{-120mm} (A)
\\
\vspace*{2mm}
\\
\includegraphics[scale=0.91]{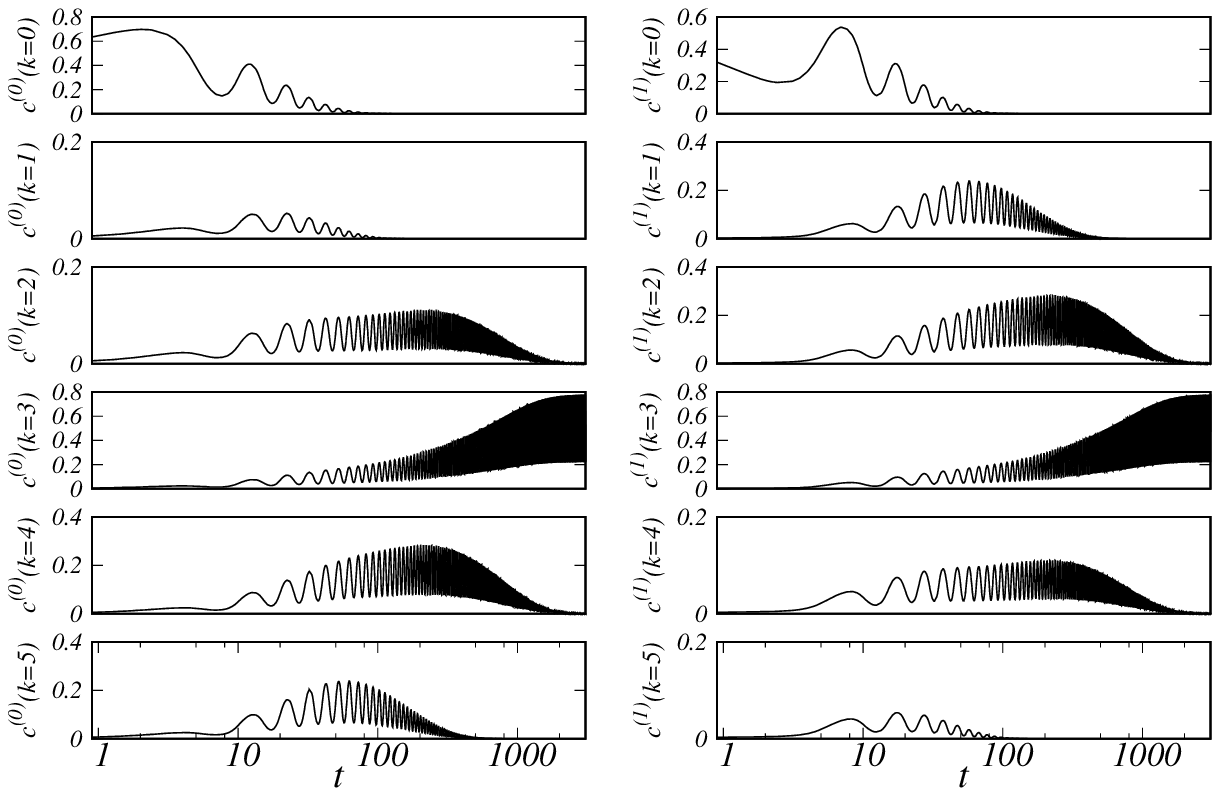}
\vspace*{-9mm}
\\ 
\hspace*{-120mm} (B)
\end{tabular}
\caption{Panel (A): Schematic diagram of the 
quasispecies model ($n_s=6$)
introduced in Section \ref{sec:Quasi}.  
Special emphasis is placed 
on the reversibility imposed by $\varphi$.
Panel (B): The time dependence 
of $n_s=6$ species fractions
in periodic environment ($T=10$). 
The calculations have been performed for the parameters  
$\Delta r=0.4$, $\mu=0.01$,  $m=0.05$.
The highest (effective) replication rate observed 
for the highest reversibility 
corresponding to  
$\varphi(3,n_s)=1/2$. 
The fractions of the other 
species are suppressed in the 
long run (limit cycle regime).} 
\label{fig:F1}
\end{figure}
\begin{figure}[!htb]
\centering
\includegraphics[scale=0.98]{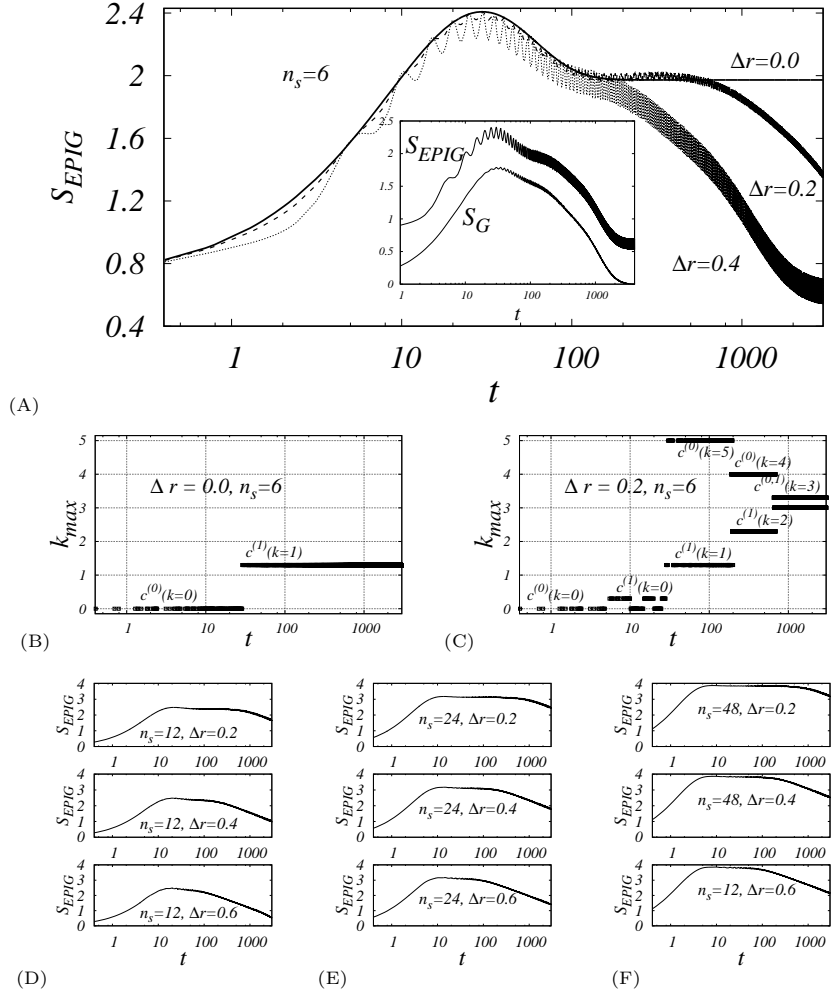} 
\caption{Panel (A) shows the time dependence of the entropy measure 
$S_{EPIG}$ (the inset includes comparison with $S_{G}$).
The difference between the sub-threshold 
($\Delta r=0$, absence of the variation) 
and super-threshold fitness environmental variations 
(alternatives $\Delta r=0.2$, $0.4$).
Calculated for the period $T=10$.  
Remaining undetermined parameters are 
$\mu=0.01$ and $m=0.05$.  
In the panels (B) and (C) we plot the time dependencies 
of the effective index 
$k_{max}(t) + z_{max}(t)  \times 0.3$. 
Here symbol $max$ is related to the species of the 
maximal abundance and the auxiliary 
constant $0.3$ is chosen for the visualization 
purposes only. For example 
in the case $z_{max}=0$ 
the plot hits the mesh line 
[see $c^{(0)}(5)$, $c^{(0)}(4), 
c^{(0)}(3), c^{(0)}(0)$ in the panel (C)]  
corresponding to the species of the maximal 
instant concentration showing the enhanced 
selection for the phenotype with highest 
reversibility, i.e. for 
$\varphi=1/2$, 
$\varphi/(1-\varphi)=1$. 
In panel (B), the evolution in the static environment remains frozen at 
$k_{max}=1$, $z_{max}=1$, 
whereas the evolution obtained for
$\Delta r=0.2$ in panel (C) 
asymptotically supports the selection 
for the phenotype with 
highest degree of reversibility 
$k_{max}=3$,  
$z_{max}=0$, $1$ and $\varphi(3)=1/2$.
(The results obtained 
for higher $\Delta r$ are qualitatively similar.) 
Panels (D), (E), (F) depict effects 
of the number of clones 
$n_s  \in \{12,24,48\}$ and 
$\Delta r\in \{0.2, 0.4, 0.6\}$ on the $S_{EPIG}(t)$.} 
\label{fig:F2}
\end{figure}

\begin{multicols}{2}

\subsection{Effective replication rate \\ of the clone} 

To investigate the behavior of the system 
of equations (\ref{eq:dczk0}) and (\ref{eq:dczk1}) we proposed an alternative effective 
description. Within this, 
the effective dynamics of total fraction $c(k,t)$ (see Eq.(\ref{eq:ctotal})) 
can be expressed by simple formula 
\begin{equation}
\frac{d c(k,t)}{dt} = ( r_{eff}(k,t) - \Phi(t))\, c(k,t) \,.
\label{eq:dceff} 
\end{equation}
Here, the effective replication rate of the $k-$th clone $r_{eff}(k,t)$ 
plays a key role for many results to follow. The phenomenology avoids 
the formal use of mutations, switching or $c^{(0)}(k,t)$, 
$c^{(1)}(k,t)$.  
The summation of the equations Eq.(\ref{eq:dceff}) 
leads to the formula  
\begin{equation} 
\Phi(t) =  \sum_{k=0}^{n_s-1} r_{eff}(k,t)  \,,
\label{eq:Phi1}
\end{equation}
which implies the interpretation of $\Phi$ as a total replication rate.   
Moreover, the consistence of Eq.(\ref{eq:Phi1}) and Eq.(\ref{eq:Phi}) 
can be accomplished for the 
"mixed-type" solution (see also \cite{Frey2017})
\begin{equation}
r_{eff}(k,t) = r^{(0)}(t) c^{(0)}(k,t) + r^{(1)}(t) c^{(1)}(k,t) \,.
\label{eq:reff} 
\end{equation}
\end{multicols}

\begin{figure}[!htb]
\centering
\includegraphics[scale=0.98]{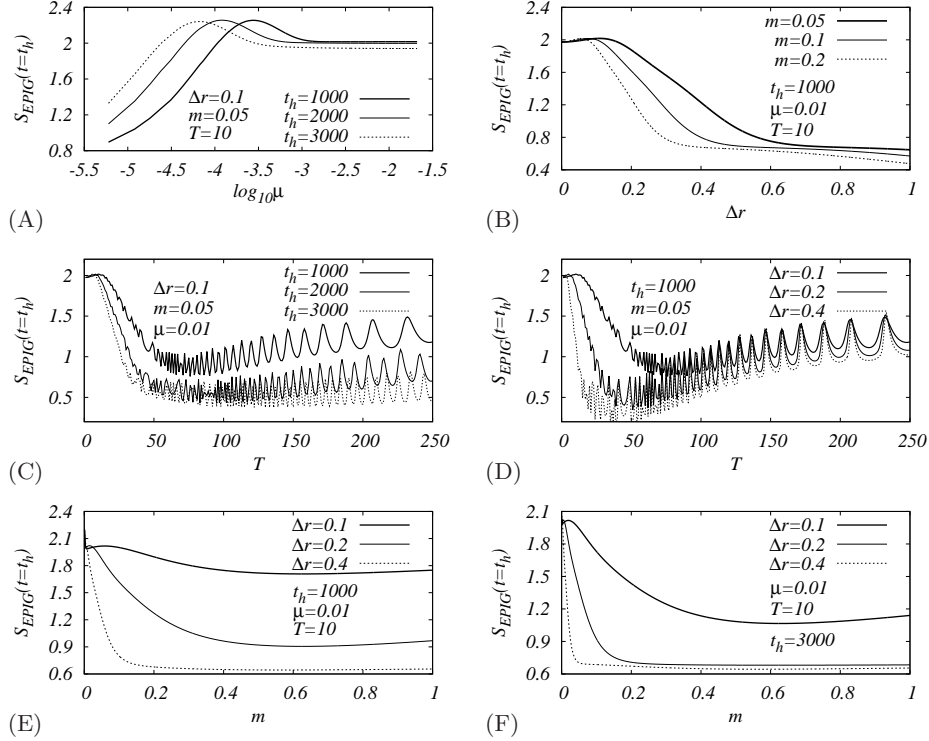} 
\caption{The differences in the efficiency of selecting extreme reversibility 
within the frame of nonequilibrium evolutionary dynamics.  
The panels show dependencies of the entropy measure 
$S_{EPIG}(t=t_h)$ calculated for various combinations of 
parameters and time horizons $t_h\in \{1000, 2000, 3000\}$.  
Panel (A) involves the threshold dependencies 
on the mutation rate $\mu$. 
Note that sufficiently large $\mu$ is needed to overcome the initial barrier 
of forming heterogeneous state from which selection can be realized. 
According to panel (B) the more intense environmental 
changes, the faster selection for the most reversible clone. 
Panels (C), (D) indicate non-monotonous dependence on $T$. 
In case (D), obtained for $\Delta r\in \{0.1, 0.2, 0.4\}$,  
the efficiency of selection reduces for the wide range of $T$ until global 
minima $T\sim (25 - 90)$ are attained. Obviously, as shown in panels (E), (F), 
selection mechanism 
can be easily accelerated 
by increasing the switching 
rate parameter $m$.}
\label{fig:F3} 
\end{figure} 

\begin{multicols}{2}
\subsection{Numerical investigation}\label{sec:Num}

Apart from a few numerical exceptions, listed explicitly in the
figure captions, the simulations have been performed for $n_s=6$ species
and the parameters $T=10$, $r_B=1,$  $\Delta r \in \{ 0.1, 0.2, 0.4\}$, 
$m=0.05,$ $\mu=0.01$. 
In agreement with intuitively expected behavior of the model, 
Fig.~\ref{fig:F1} shows gradual qualitative progress 
in the redistribution of the species fractions.  
At the intermediate time scales ($10 \,a.u.-100\, a.u.$),
there is a period of "hesitancy", 
where all the clones  
$k>1$ nearly counterbalance their
respective competitors. 
During this short initial phase, 
the variable environment causes that  
the putative equality 
among the clones $k=1,2,\ldots, 6$ 
gradually vanishes and the highest reversibility 
due to $\varphi = 1/2 $ corresponding to the ratio 
$\varphi/( 1-\varphi) 
= 1$ attributed to the clone 
$k_{\rm central} = n_s/2  = 3$, leads at the 
long run ($\geq 1000$) 
to the largest $r_{eff}$ [see the 
analytical approximation given by 
Eq.(\ref{eq:rBb}) in the Appendix \ref{Ap:Long}].

Fig.\ref{fig:F2} shows the non-trivial 
transient states and the initial increase 
common for $S_{EPIG}$ and $S_{G}$. 
The largest differences between
$S_{EPIG}$ and $S_{G}$ are localized 
mainly in the long-time asymptotics.  
Here, the main question arises to what extent environmental 
variability affects the long-term heterogeneity 
of the populations. There are many indicia that populations
obtained for large and small $\Delta r$ significantly differ.
Fig.\ref{fig:F2} depicts both the entropy measures as functions of $\Delta r$, which is the main determinant of environmental 
variability. The variability causes the convergence of $S_{EPIG}$
to the positive limit value, while trivial limit $S_G\rightarrow 0$ indicates 
dominance of the single species.
For small $\Delta r$, there are only very low evolutionary benefits 
due to switching between phenotypic states, which leads to extremely slow evolution. 
This means that numerical approach becomes 
inappropriate to capture the subtle 
differences between non-equilibrium decay and long-time limit cycle behavior.  

The numerical problems 
stimulated development of analytical 
approximation discussed in Appendix \ref{Ap:Long}. 
Both approaches are consistent in indicating 
of the fact that convergence to the most 
reversible clone is faster as 
$ \Delta r$ increases 
[see panel (C) of Fig.\ref{fig:F2}\,]. 
In addition, it appears that for given time 
horizon $t_h$ the increase 
in $\Delta r$ leads to the threshold-like behavior 
[see Fig.\ref{fig:F3} (B)]. According 
to Fig.\ref{fig:F2} (B), 
without sufficient environmental 
variability most of the 
population remains trapped in the clone $k = 1$.
Fig.\ref{fig:F3} summarizes the entropic responses 
induced by $\mu$, $\Delta r$, $T$ and $m$. 
For example, panels (C) and (D) show 
how the choice of $T$ affects the efficiency 
of the selection of the most reversible strategy.   
It turns out that large enough $T$ supports 
rapid selection. In agreement with the intuitive expectations, 
the process can be reinforced 
by super-threshold value of $m$.  

\end{multicols}

\begin{table}[bht]
\begin{center}
{\small
\begin{tabular}{|ccccc|c|} 
\hline
                        &
                        &    
 $\downarrow\, \mu$     & 
 $\downarrow\, \mu$     & 
 $\downarrow\, \mu$     & 
\fbox{{\scriptsize $C_P$\,\,} $\leadsto $
{\scriptsize mutations}}
 \\
                    &
                    &     
$C^{(0)}_{L0} $     &  
$C^{(0)}_{L1} $     &   
$C^{(0)}_H    $     &  
\\
                    &
 \hspace*{4mm} $\nearrow$         &    
 \fbox{$\alpha_H$ $\big|$ $x_p$}           & 
 \fbox{$\alpha_L$ $\big|$ $1/x_p$}         & 
 \fbox{$\alpha_H$ $\big|$ $x_p$}             &
\fbox{
 \begin{tabular}{l}
{\scriptsize drug}\\
{\scriptsize efficiency}  
\end{tabular} $\big|$ {\scriptsize specificity}} 
 \\
 $C_P \longrightarrow$   &
 {\large $6 \mu$}         &    
 \fbox{$\varphi_L \, \updownarrow $\, $m$}    & 
 \fbox{$\varphi_L \, \updownarrow $\, $m$}    &    
 \fbox{$1/2\,\, \updownarrow$\, $m$}          & 
\fbox{{\scriptsize commitment} \, \,${\big|}$ {\scriptsize reversibility}}  
\\
                           &
\hspace*{4mm}$\searrow $                &    
\fbox{$\alpha_L $ $\big|$ $1/x_p$}         & 
\fbox{$\alpha_H $ $\big|$ $x_p$}         &    
\fbox{$\alpha_L$ $\big|$ $1/x_p$}              &  
\fbox{
\begin{tabular}{l}
{\scriptsize drug}\\
{\scriptsize efficiency}  
\end{tabular} $\big|$ {\scriptsize specificity}}      
\\                       
                    &
                    &  
$C^{(1)}_{L0}$      &  
$C^{(1)}_{L1}$      & 
$C^{(1)}_H$          &  
\\
                    &    
                    & 
 $\uparrow\,\mu$    &    
 $\uparrow\,\mu$    & 
 $\uparrow\,\mu$    &
\fbox{
{\scriptsize $C_P$\,\,}$\leadsto$ {\scriptsize mutations}}
 \\
\hline
\end{tabular}}
\end{center}
\caption{The scheme depicts the relations between 
abundances $C_P$,  
$C^{(z)}_{L0}$, $C^{(z)}_{L1}$, $C^{(z)}_H$; $z\in \{0,1\}$
mutation ($\sim\mu$) and switching ($\sim m$) processes; 
the variants of phenotypic state commitments 
$\{\varphi_L$, $1/2\}$ [see Eq.(\ref{eq:JL0})],
drug efficiencies $\{\alpha_L$, 
$\alpha_H$, $1\}$ [see Eq.(\ref{eq:alpha00})],
and drug specificities $\{\,x_p$, $1/x_p\}$ 
[see Eq.(\ref{eq:aHxp})] within 
the model represented 
by Eq.(\ref{eq:dCPdt}).}
\label{table:tab1}
\end{table}

\hrule
\begin{multicols}{2}
\section{Model of immunotherapy}\label{sec:Evo}

In the next we apply the above framework to
investigate evolutionary aspects of the 
resistance to immunotherapeutic drugs caused by 
phenotypic reversibility. We built the new model 
using the exogenous simplified variant of the Kirschner-Panetta (KP) model of immunotherapy \cite{Kirschner1998}. 
The original KP model consists of the three autonomous equations,
for the effector cells production, tumor growth with tumor clearance,
and, finally, equation for cytokine interleukin-2 (IL-2) production. 
Essential in the original KP model is the boosting of the IL-2 
against tumors. There are many versions of the model that include
simplified but biologically plausible approaches  \cite{Talkington2017}.
In \cite{Tsygvintsev2013} the gene therapy characterized by populations
of effector and cancer cells was described by means of two coupled 
autonomous ODE. Recently, two-populations cancer model \cite{Piretto2018}
with competitive interaction has been used to investigate  
the combination of immunotherapy and chemotherapy. 

Here, we present a simplified model of a single cancer clone where 
a population of cancer cells varies accordingly to exogenous ODE
\begin{equation} 
\frac{d C}{dt}  = r C (1- C) -  a  \frac{e(t) C}{C+ g} \,. 
\label{eq:rCae}
\end{equation}
Consistently with the original KP model, the efficiency of therapeutic 
interventions is described by the clearance parameter $a$.
The parameter $g$ plays the role analogous to that in the Michaelis-Menten
kinetics. Regarding the treatment classification based on the  
rate of tumor cells killing, the model term $-a e(t) C/(C+ g)$ 
corresponds to the cytotoxic activity of non-logkill type \cite{Onofrio2012}. 
Unlike the quasispecies model, the rate of replication $r$ does not change.  
The classical logistic term 
$r C (1-C)$ is used for the nutrient-limited 
cancer growth of the rescaled abundance of tumoral cells $C$. 
Its use [loosely analogous to $\Phi$ from Eq.(\ref{eq:Phi1})] 
causes that tumor size stabilizes at $C\leq 1$.
(The carrying capacity is equal to $1$ in here adopted scaling). 

Dynamics of environment is mediated by the population of the effector cells. 
We assume that population size $e(t)$ 
varies according formula 
\begin{equation}
e(t) = e_B + \Delta e \cos\left(\frac{2\pi t}{T} \right) \,.
\label{eq:eBt}
\end{equation}
This simplification exploits a few aspects:
(i)~the exogenous formulation is parametrically less demanding 
than the original autonomous (endogenous) KP model; 
(ii)~it can be easily linked to the harmonic model of 
replication rate $r^{(z)}(t)$ proposed in Eq.(\ref{eq:r01}) 
with the exception that it cancels the dependence on 
$z\in \{0,1\}$ (the phenotypic state 
dependence is restored in the 
Subsection \ref{sub:divim} by introducing additional parameters 
to represent therapeutic interventions); 
(iii)~it avoids the ODE stiffness problems typical of the original KP formulation; 
(iv)~the harmonic character of 
Eq.(\ref{eq:eBt}) is qualitatively consistent 
with KP phase diagrams \cite{Kirschner1998}
belonging to the dynamical regimes with 
some {\em limit cycle attractors};
(v)~periodicity is consistent with the therapeutic option 
for immuno-oncological dynamics \cite{Onofrio2010}.

Before studying more complex 
numerical examples, we continue with the parametrization of the elementary model 
based on Eq.(\ref{eq:rCae}) and Eq.(\ref{eq:eBt}). 
According to the results, the system preliminary simulated 
with the initial condition $C(0)=0.1$ 
for ten periods exhibits 
the largest susceptibility to the changes in $g$ for the parameters 
\begin{eqnarray}
\label{eq:parame} 
\begin{tabular}{lll}
$  e_B = 0.5$\,,       &              
$  \Delta e = 0.4$\,   &  \hspace*{-5mm}
\,\,\,(\mbox{\small analogs \,\,of} $ r_B$ \mbox{\small and} $\Delta r$)\,,   
\\
$  g=0.45$\,,  &    
$  \,a = 1$\,,     &  \hspace*{-4mm} 
$ \, \,T=10$\,,\,\,\,       
$  r=1$ \,(\mbox{\small fixed})\,.
\end{tabular}
\end{eqnarray}
The parameters 
are intended for further simulations of 
multiclonal populations of cancer cells. 

\subsection{Diversification of the effects of immunotherapy}\label{sub:divim}

To illustrate the role of reversibility of phenotypic switching 
in the evolution of the therapy resistance,
we construct the evolutionary model where
reversibility is considered along with intratumor heterogeneity 
reflected by the clone-dependent and state-specific sensitivity to drug.
It works with the population of seven rescaled  instant species abundances, 
forming the set $ set_E(t) \equiv \{C_P(t)$, $C^{(0)}_{L0}(t),$ $C^{(1)}_{L0}(t),
C^{(0)}_{L1}(t), C^{(1)}_{L1}(t), $ $C^{(0)}_{H}(t), $
$C^{(1)}_H(t)\}$, where $C_P(t)$ denotes the
abundance of the primary cancer cells.

Four abundances $C^{(z)}_{L0}(t)$, $C^{(z)}_{L1}$, $z\in \{0,1\}$
correspond to two clones $L0, L1$ with low reversibility (indexed as $L$) between states.
The clones are characterized
by the commitment $\varphi_L\in (0,1/2 \rangle$.
Here, the a priori tendency toward low reversibility 
(with the obvious exception for $\varphi_L = 1/2$)
is ensured by the combination of "asymmetric" 
commitments $\varphi_L$, $1-\varphi_L$
(which occur in the complementary pairs of the switching flows).
The supplementary lower numerical 
index $0$ in $C^{(z)}_{L0}$ (or $1$, respectively)
indicates that immunotherapy is specially designed to have more profound
impact on the phenotypic state $0$ (or $1$) (therapeutic effectiveness
is below denoted as $\alpha_H$), whereas the alternate phenotypic state
is eliminated with lower efficiency 
($\alpha_L$, see below).
The commitment equal to $1/2$ (high reversibility) is imposed on
the clone $H$ characterized by the abundances 
$C^{(0)}_{H}$, $C^{(1)}_H$.

The overview of the theoretical structure of the model is schematically highlighted 
in Table \ref{table:tab1}. 
In contrast to the quasispecies model, the sum  
\begin{equation}
C_{\Sigma} \equiv C_P+
 \sum_{z=0,1}\,
\left(C^{(z)}_{L0}+C^{(z)}_{L1}+C^{(z)}_H\right)
\label{eq:CSig}
\end{equation}
is not normalized to unity but, instead, 
$(1-C_{\Sigma})$ is used to construct the logistic form  
$r C (1- C_{\Sigma})$ 
which guarantees $C_{\Sigma}(t)\leq 1$ for proper 
initial conditions.  
Similar term that sets 
upper limit on the abundance 
of the mixed populations 
can be found elsewhere \cite{Gatenby2003V,Gerlee2015,Piretto2018}.

As we pay attention to the problem of reversibility within
the context of heterogeneity, 
the switching 
flows within the clones $L0$, $L1$, and $H$  
are defined as it follows  
\begin{eqnarray}
J_{swL0} &=& 
(1-\varphi_L) C^{(1)}_{L0} - 
\varphi_L C^{(0)}_{L0}\,,
\label{eq:JL0}
\\
J_{swL1} &=& (1-\varphi_L) 
C^{(1)}_{L1} - 
\varphi_L 
C^{(0)}_{L1}\,,
\nonumber
\\
J_{swH} &=&
\frac{1}{2}
 \left( 
C_H^{(1)} - 
C_H^{(0)} \right) \,.
\nonumber 
\end{eqnarray}
All population rates are defined by the universal formula 
consisting of the adaptations of Eq.(\ref{eq:rCae}) and 
former variant of the switching flow model
\begin{eqnarray}
&& {\mathcal R}(\tilde{C}, \tilde{J},
\tilde{\alpha}, N_{\mu} )
= r \tilde{C} ( 
1 - C_{\Sigma})\,\,\,\,\,\,\,\,\,
\hspace*{12mm}
\nonumber
\\
\label{eq:Rform}
\\
&&
\hspace*{10mm} +
m \tilde{J} - \tilde{\alpha} a \,
\frac{e(t) \tilde{C}}{
C_{\Sigma}+g} 
+ N_{\mu} \,\mu C_P\,,
\nonumber 
\end{eqnarray}
where 
$\tilde{C}, \tilde{J},$
$\tilde{\alpha}, N_{\mu}$ are some auxiliary
general variables which are substituted by
some specific variables and constants: 
$\tilde{C} \in set_{E}$, 
$\tilde{J} \in  \{\, J_{swL0}, J_{swL1}, J_{swH} \} $,
$\tilde{\alpha} =\{\,\alpha_L, \alpha_H\,\}$, 
$N_{\mu}\in \{\,-6,\,+1\}$. 
In comparison with Eq.(\ref{eq:rCae}), 
we also include mutation mechanism 
($\sim N_{\mu}\mu C_P$)
and switching ($\sim m \tilde{J}$)  
already used in the quasispecies simulations.   

The contributions related to the switching process are 
adopted from \cite{Fedotov2007}. The alternatives
$J_{swL0}$, $J_{swL1}$, $J_{swH}$ (might loosely correspond to
"escape routes" for the respective cancer sub-populations \cite{Kemper2014}. 
Conversions between drug-resistant and drug-sensitive clones 
are the main issue of the resulting 
construction considered in line with the abstract 
model \cite{Liao2012} used as a prototype. 
To study bacterial persistence under conditions of  antibiotic stress, a similar,
two-component population model was proposed in \cite{Kussell2005a}.
Finally, the technical details of simultaneous incorporation 
of mutations and the switching processes
into the model of epigenetically structured population 
can be found elsewhere \cite{Libby2011}. 

To introduce the clone- and state-dependent sensitivity to drug
into the model, the scalar clearance parameter $a$ in Eq.~(\ref{eq:rCae}) 
is in Eq.(\ref{eq:Rform}) replaced by the species-dependent
term $\alpha_j^{(z)} a$, $j\in \{0,1\}$, $z\in \{0,1\}$ 
with the prefactor representing a modifier of the drug efficiency
of the phenotype $z$ of the clone $j$. 
To stay consistent with the formal structure of the model, $\alpha_j^{(z)}$
is represented by four matrix elements, $\alpha_0^{(0)}, \alpha_0^{(1)},
\alpha_1^{(0)}, \alpha_1^{(1)}$, related to the particular clone and phenotype.
Nevertheless, to keep the conceptual model instructive, we assume
only two basic levels in the matrix elements, one for
diagonal, one for non-diagonal elements, respectively
\begin{eqnarray} 
\left(
\begin{array}{ll}
\alpha_{0}^{(0)}  &  \alpha_{1}^{(0)}   \\   
\alpha_{0}^{(1)}  &  \alpha_{1}^{(1)}   
\end{array}
\right)  
= \left(
\begin{array}{ll}
\alpha_{H}   &   \alpha_{L}  
\\ 
\alpha_L   &  \alpha_H
\end{array}   \right)\,.
\label{eq:alpha00}
\end{eqnarray}
We note that diversity of the forms of $ \mathcal{R} $ conditioning 
diversity within $set_E$ stays sufficient as it is not based explicitly
on the effect of  $\tilde{\alpha}\in \{ \alpha_L, \alpha_H\}$, 
but on the possible combinations of $\tilde{\alpha}$ 
with the commitment terms $ \varphi_L, 1 - \varphi_L, \frac{1}{2} $ 
coming from $ J_{swL0} $, $J_{swL1}$
and $ J_H $ as well.

Both $\alpha_L$, $\alpha_H$, representing low/high therapeutic
effects, respectively, 
can be further parametrized by the 
single auxiliary parameter  $x_p\geq 1$
\begin{eqnarray}
\alpha_H  (x_p)
&=& 
x_p \,, \qquad 
\alpha_L (x_p) = \frac{1}{x_p}\,.
\label{eq:aHxp}
\end{eqnarray}
The parametrization is designed to satisfy requirements
$\alpha_{0}^{(z)} \alpha_{1}^{(z)}  
=  \alpha_{L}(x_p) \alpha_H(x_p) = 1$
and $0<\alpha_L\leq 1 \leq \alpha_H$ 
for both phenotypic states 
$z\in \{0,1\}$. In the special boundary case $x_p=1$ which implies 
$\alpha_L(1)=\alpha_H(1)=1$, there is no modification of $a$.  
To sum up, the parameter $x_p$ can be interpreted
as a measure of {\em specificity} of {\em immunotherapeutic intervention}. 
To present a microscopic view 
of the problem studied, we outline an 
interpretation in which $x_p$ integrates the basic mechanisms, 
such as the specific efficacy of the tumor-associated 
antigens, the ability of 
tumors to activate T-cell responses 
including the tumor recognition and the 
induction of cancer cell deaths \cite{Ilyas2015}.

By combining Eq.(\ref{eq:JL0}) with Eq.(\ref{eq:Rform}),  the population dynamics of  
$(1+6)$ tumor species can 
be expressed as it follows  
\begin{eqnarray}
\frac{d C_P}{dt} & = & {\mathcal R}(C_P, 0, 1, -6)  \,,
\label{eq:dCPdt}
\\
\frac{d C^{(z)}_{L0}}{dt} 
&=&
{\mathcal R}(C^{(z)}_{L0},
(-1)^z J_{swL0},
\alpha_{0}^{(z)}, +1)\,,
\nonumber 
\\
\frac{d C^{(z)}_{L1}}{dt} 
&=& 
{\mathcal R}(C^{(z)}_{L1},(-1)^z J_{swL1},\alpha_{1}^{(z)}, +1)\,,
\nonumber 
\\
\frac{{d C}_H^{(z)}}{dt} &=& 
{\mathcal R}(C_H^{(z)},(-1)^z J_{swH} ,
\alpha_0^{(z)}, +1) \,.
\nonumber 
\end{eqnarray}
The first formula 
of the list 
expresses that phenotype switching 
absents (zero at the second position of ${\mathcal R}$) 
in the primary tumor clone. 
This choice is consistent 
with the quasispecies model. 
In addition, there is no modification of the drug sensitivity 
$a$  (${\tilde \alpha}=1$).  The factor $(-6)$ expresses that
mutation process modifies 
the primary tumor growth by subtracting rate factor $6 \mu C_P$. 
We assume that  the "most reversible" 
and "symmetric" transitions between phenotypes 
corresponding to $C_H^{(0)}$
and $C_H^{(1)}$  
are violated by the factor $\alpha_0^{(z)} \neq 1$. 

As in the case of quasispecies model, now we propose
measures that allow better interpretation of the results. Again, the 
appropriate transformations allow us to define proper 
measures of heterogeneity.
For all $C_E \in set_E$
the rescaling $(C_E/ C_{\Sigma})$
can be performed, that guarantees $\sum_{C_E\in set_E} (C_E/ C_{\Sigma} ) =1$.
In this case, the tumor can be monitored 
by epigenetic entropy-like measure 
\begin{eqnarray}
S^C_{EPIG}  
= -  \sum_{C_E\in set_E}
\frac{C_E}{C_{\Sigma}} 
\ln
 \frac{C_E}{C_{\Sigma}} \,. 
\end{eqnarray}
In analogy with 
Eq.(\ref{eq:ctotal}), we define four-element 
auxiliary set  
$set_{G} \equiv \{C_P,$  
$ C^{(0)}_{L0}+C^{(1)}_{L0},  $
$ C^{(0)}_{L1}+C^{(1)}_{L1},  $ 
$ C^{(0)}_{H}+C^{(1)}_{H} \}   $  
which allows modification of the original definition of 
Eq.(\ref{eq:SG}) to the form 
\begin{eqnarray}
S^C_{G} =  - \sum_{C_G\in set_{G}}
\frac{C_G}{C_{\Sigma}} 
\ln
 \frac{C_G}{C_{\Sigma}} \,. 
\label{eq:sumSCG}
\end{eqnarray}
\end{multicols}

\begin{figure}[!bhtb]
\centering 
\includegraphics[scale=0.982]{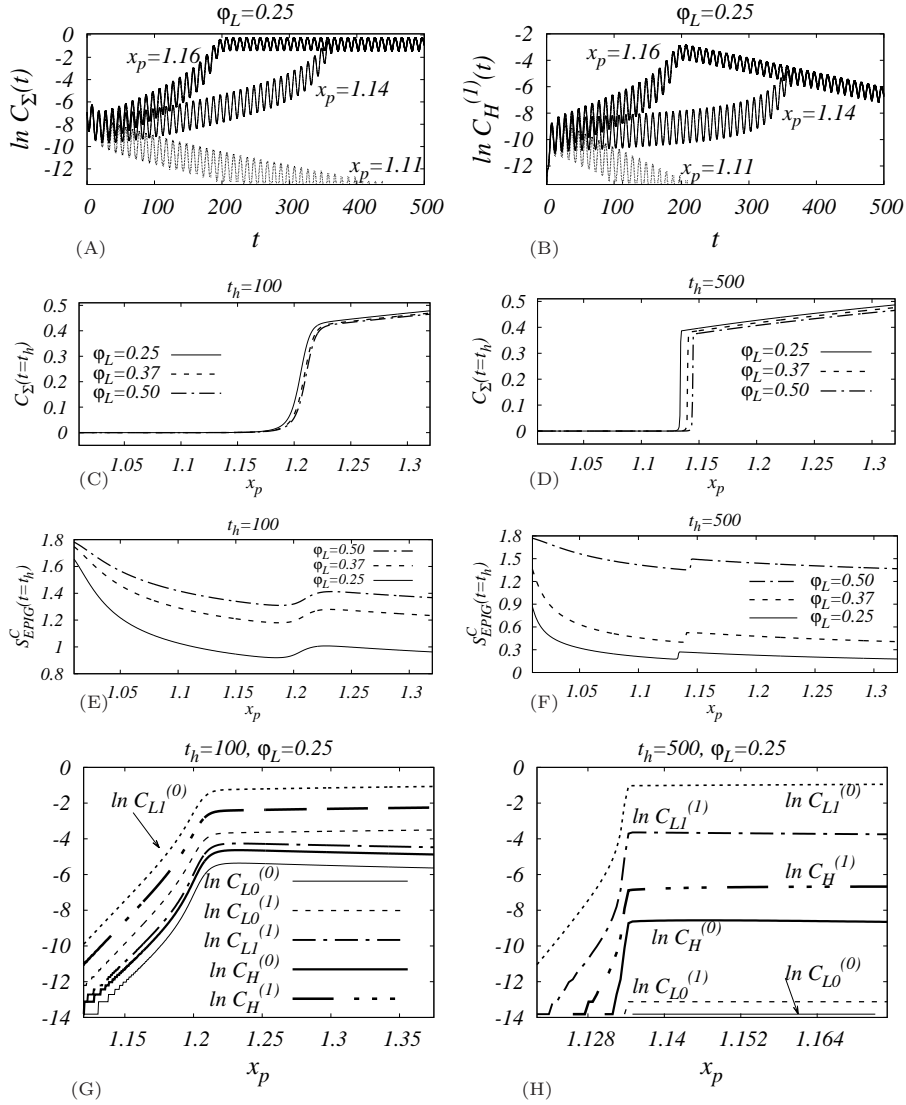}      
\caption{The numerical results 
obtained for the model 
incorporating evolution of reversibility and resistance 
to immunotherapy. 
In addition to other parameters 
we kept $\mu=0.01$, $m=0.05$ and 
$x_p$, $\varphi_L$, $t_h$ specified in the panels.
The tumor is variable in size 
($C_{\Sigma}(t)$) and phenotypic 
structure ( $C_E\in set_E$).  
Panels (A), (B) show some differences 
in the course of the tumor progression 
($x_p=1.16$, "rapid progression";  
$x_p=1.14$, "slower" progression; 
$x_p=1.11$, "shrinking")  
for three  selected values of $x_p$ 
characterizing qualitative differences in the 
therapeutic effects on the phenotypic states;
(e.g. drug specificity $x_p=1.11$ implies 
$\alpha_L\simeq 0.9009$);
Panel (B) shows specific phenotypic dynamics. 
It shows that despite the high reversibility 
[commitment $1/2$ incorporated into $J_{swH}$ 
in the case of clone $[H; z \in \{0,1\}]$],  
the abundance $C_H^{(1)}(t)$ drops  
at the late times as a result of the loss of the ability to compete. 
Panels (C), (D), $\ldots$ (G), (H)
reveal details of $x_p$  influence calculated for two time 
horizons $t_h \in \{100,\,500\} \,$. 
The threshold between slow and rapid 
tumor progression can be identified.
The threshold character is confirmed by the fact, 
that the transitions due to $x_p$ are getting sharper
as the $t_h$ increases. 
The sharpening due to a change in $t_h$ from 
$t_h=100$ to $t_h=500$ is also 
noticeable for the panels (G), (H), 
where more detailed 
information on the population structure is  available.
The panels (E), (F) depict the formation 
of the threshold values in the Shannon entropies  
as heterogeneity measures. Panels (G), (H) show that  
$C_{L1}^{(0)}$ dominates over a wide range of $x_p$ 
although the corresponding 
$\varphi_L=0.25$ is quite far from the 
ideal reversibility corresponding to $1/2$. 
From panel (H) we see that the 
significant long-term desirable effect 
of therapy persists only in the case 
of $C_{L0}^{(0)}$ and 
partly in the case of $C_{L0}^{(1)}$.}
\label{fig:F5}
\end{figure}

\begin{figure}[!thtb]
\centering 
\includegraphics[scale=0.93]{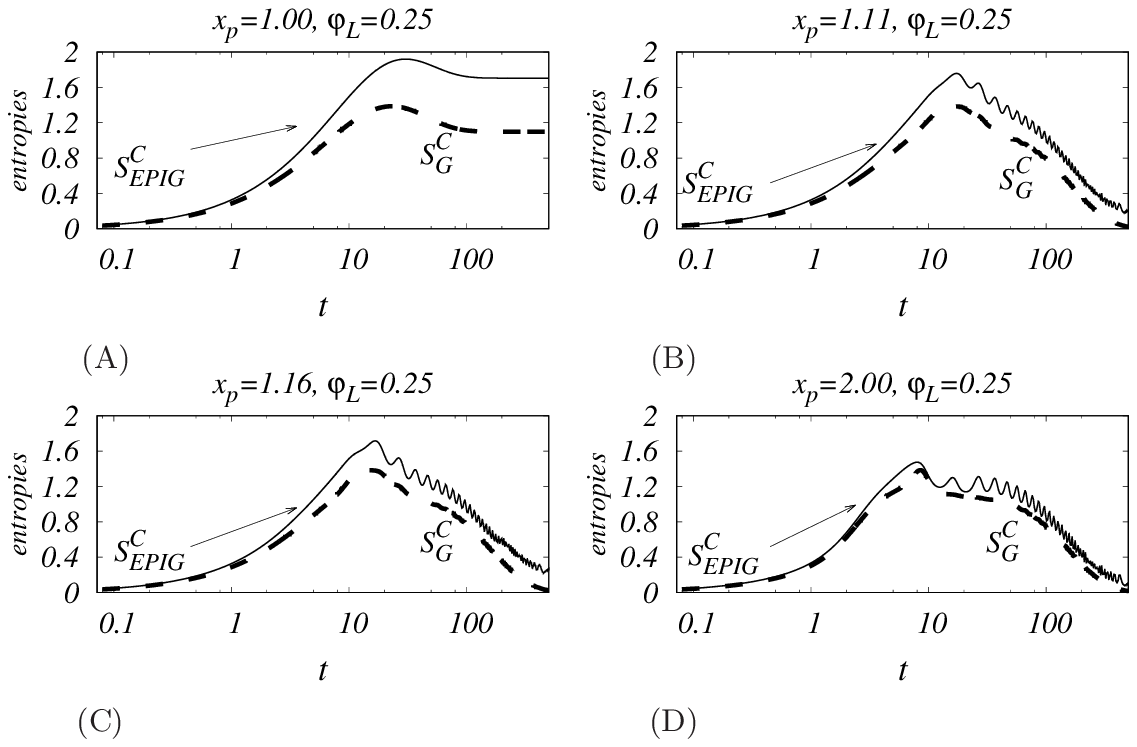}
\caption{The evolutionary dynamics reflected by the heterogeneity measures 
$S^C_G(t)$ and $S^C_{EPIG}(t)$.  
The numerical results obtained for 
the additional parameters $\mu=0.01$, $m=0.05$ 
and $x_p, \varphi_L$ listed on the top of figure panels. 
The scenarios corresponding to different 
$x_p$ are qualitatively similar:  
the initial increase associated with a large number 
of possibilities is replaced in the peak 
zones by the favorable 
growth of selected clones,  i.e.  clonal expansion.
A weakened preference due to $x_p=1$ 
in panel (A) leads to the highest entropy 
among the considered outcomes. 
Panels (B), (C) show the transition stages.  Panel (D) indicates 
that differences between epigenetic and genetic 
entropy measures are almost entirely 
neutralized if selective pressures due to immunotherapy 
are strong enough.  For (B), (C), (D) panels the after peak 
behavior of $S^C_{G}(t)$ changes 
smoothly contrary to "jagged" form 
of $S^C_{EPIG}(t)$.} 
\label{fig:F6}
\end{figure}

\hrule
\begin{multicols}{2}
\subsection{Simulation results - evolution of the therapeutic resistance}\label{sub:SimRes}

The results of numerical simulations are depicted 
in Fig.\ref{fig:F5} and Fig.\ref{fig:F6}. They are obtained for the initial conditions  
\begin{eqnarray}
C_P(0)  &=& C_{P0}\,,
\label{eq:InitC}
\\ 
C^{(z)}_{L0}(0) &=& C^{(z)}_{L1}(0)=C^{(z)}_{H}(0) = 0\,, 
\,\,\,\, 
z \in \{0,1\} \nonumber\,.
\end{eqnarray}

\vspace*{2mm}

\vspace*{2mm}

\vspace*{2mm}

\vspace*{2mm}

\pagebreak 

In the following calculations the parametric value $C_{P0}=0.001$
and parameters defined by Eq.(\ref{eq:parame}) from the Section
\ref{sec:Evo} are used, with the exception of $x_p$ and $\varphi_L$,
which are listed in the respective figure captions.
The applied initial conditions resemble 
those considered in the work \cite{Lorz2015}.

As is clear from the respective panels of Fig.\ref{fig:F5}, the different combinations of 
the drug specificity parameter $x_p$ 
and phenotypic commitment parameter 
$\varphi_L \in \{0.25, 0.37, 0.5\}$ 
trigger different competitive 
behaviors, which are manifested 
for different time horizons 
$t_h \in \{\,100, \, 500\,\}$. 

A more detailed look in the panels (A), (B) of Fig.\ref{fig:F5}
which show the monitoring of $C_{\Sigma}$ and $C^{(1)}_H$
reveals additional systemic features.
For such characteristics, quite strong preference
defined by $x_p = 1.16$ implies very 
rapid growth, with as few as $200$ time units 
nearly sufficient to achieve  
carrying capacity. Although the abundance $C_{\Sigma}$ stabilizes
at its upper limit, there are still internal changes in the clonal subpopulations
which are visible in variations of $C^{(1)}_H$.
The competition under saturated conditions 
leads to the loss of $C^{(1)}_H$ dominance, giving
advantage to other clones. 
Owing to locating simulation intentionally in the area with
a relatively high sensitivity to $x_p$, even a small reduction
to $x_p = 1.14$ makes anticancer effect more apparent.
In such case, $300$ time units are necessary to attain saturation. 
This slowing down of the progression relates to the fact 
that more distributed therapy weakens not only dominant 
but, at least partially, other phenotypic variants as well.
The dynamics presented in the panels (A), (B) 
enabled us to focus on the appropriate time horizons 
$t=t_h$ for which, consequently, dependence of 
the respective values $C_{\Sigma}(t=t_h)$ on $x_p$ was calculated. 
As the panels (C), (D) show, 
larger $t_h$ leads to more pronounced transition in $C_{\Sigma}$.
The entropy dependencies depicted in panels (E), (F) 
confirm the global trend which can be attributed to the enhanced 
selection due to higher $x_p$.  
We note, that this result was achieved presuming that all tumor clones are well identifiable
which, however, contradicts to currently known limitations 
of biopsy and histology \cite{Gerlinger2012}.

However, the slight global decrease in $x_p$ is not the only remarkable 
trend of the entropy as, apart from it, the local modulation caused by the 
presence of the threshold is present as well. For a more comprehensive 
understanding of the entropic dynamics, 
Fig.\ref{fig:F6} is presented, 
since projection of entropy values extracted 
exclusively for $t=t_h$ is not very illustrative 
Besides the plots of $S^C_G(t=t_h)$ and $S^C_{EPIG}(t=t_h)$, 
Fig.\ref{fig:F6} highlights the fact that entropy scenarios 
are universal owing to their similarity 
with those depicted in Fig.\ref{fig:F2}. 

Interestingly, regarding the time dependencies $S^C_G(t)$ and
$S^C_{EPIG}(t)$, both models, the quasispecies as well as the cancer population
model, reveal many common and universal features, such as the overall
single-peaked shape, direct proportions $S_{EPIG}\sim S_G$ 
and undulating $S^C_{EPIG}(t)$, which contrasts with much smoother $S^C_G(t)$.
In addition, the findings imply that the amount
of information/heterogeneity expressed by $S^C_{EPIG}$ exceeds 
$S^C_G$, which is consistent with the assumption 
of the additional epigenetic information \cite{Bjornsson2004},
which warns that a "summative" framework 
[here represented by Eq.(\ref{eq:sumSCG})] 
can overlook certain short-term details of the phenotypic population structure. 
Similarly as in the case of quasispecies model, the threshold effect emerges.
In this case, its importance derives from its relevance 
to the therapeutic efficiency since the threshold separates 
growth regime from the regime of the slow growth
or decay dynamics. Further details related to the investigation of the threshold 
sensitivity to the phase differences of the environment as well as supplementary
algebraic characteristics derived for ODE system Eq.(\ref{eq:dCPdt})
are listed in the Appendix \ref{Ap:Thresh}.

Fig.\ref{fig:F5}(H) displays the formation of the asymptotic regime 
in the region above the threshold. As one can see, the following 
ascending ordering of the abundances 
stabilizes at long times ($t_h\geq 500$):
\begin{equation}
C^{(0)}_{L0} <   C^{(1)}_{L0} <  
C^{(0)}_{H}  <   C^{(1)}_{H}  <  
C^{(1)}_{L1} <   C^{(0)}_{L1}\,.
\end{equation}
The ordering at the level of the pairs   
$ C^{(0)}_{L0} < C^{(1)}_{L0} $,
$ C^{(0)}_{H}  < C^{(1)}_{H}   $,
$ C^{(1)}_{L1} < C^{(0)}_{L1} $ 
is the consequence of imposed 
immunotherapeutic effects represented by 
$\alpha_L  <  \alpha_H$ 
involved in Eq.(\ref{eq:Rform}). 
Although design of $J_ {swH}$ is originally motivated 
to demonstrate superiority of reversible clones in the variable 
environments,  our results show that pathway to drug resistance 
is not necessarily associated with the 
selective advantage of the clone $H$ (with the 
structure $[H; z\in \{0,1\}]$). 
The situation, in which the abundances corresponding to $L1$ 
exceed those of $H$ (which are more populated than $L0$)
is implied by the role 
of $\varphi_L$ and 
$\alpha_H$ (or $\alpha_L$) in variable environment.
A qualitative explanation has origin in the structure of 
$J_{swL0}$ [see Eq.(\ref{eq:JL0})].  
According to it, the choice $\varphi_L\leq 1/2$ 
favors the phenotypic state 
$[L0; z = 0]$ compared to $[L0; z = 1]$. 
Similarly,  due to net effect of $\varphi_L$, 
the structure of $J_{swL1}$ suggests commitment to  
$[L1;\, z = 0]$ thereby providing higher fitness than 
$[L1;\, z = 1]$. But the immunotherapeutic effect of
$\alpha_0^{(0)} = \alpha_H $ is strong enough 
to cause a significant reduction of 
$C_{L0}^{(0)}$. However, through evolution, the proliferative 
potential is redirected toward $C_{L1}^{(0)}$, i.e. $[L1; z = 0]$, 
where cancer encounters 
only weakened immune reaction represented by 
$\alpha_1^{(0)}=\alpha_L$. 

These qualitative explanations reflect complexity of the resistance-relevant processes 
and emphasize importance of computational effort to better understand 
their role in immunotherapeutic procedure. As a consequence,  a single causal 
relationship between effects and the respective specific factor is not sufficient
for complete understanding, as there is no single factor leading 
to the corresponding effect. The results show that the 
selective phenotypic advantage is formed due to a combined 
parametric effect comprising cellular response
to drug specificity, phenotypic commitment, environmental
dynamics, as well as other mechanisms 
included in the model.

\section{Discussion}

In the paper, we have proposed a minimalist concept of nonequilibrium modeling
of the evolution of phenotypic reversibility in time-varying environments.
The results demonstrate, at the methodological level, that the proposed
ODE models can be helpful in understanding of the characteristic
features of the evolutionary dynamics of genetic and epigenetic
phenotypic states. 
At some stylized level, the space of alternatives is well 
described by the presented quasispecies and population models.
As we have shown, a possible evolutionary process can 
be better understood when the structure of the underlying model
involves combinations of low and high parametric values.
In particular, the former of the two studied models, the quasispecies model,
confirmed that under specific symmetric environmental conditions 
[where according to  Eq.(\ref{eq:r01}) the replication rates 
of phenotypic states move in symmetrically opposite 
directions $d r^{(1)}(t) = - d r^{(0)}(t)$] 
the clones with higher reversibility 
may outcompete their less reversible
rivals, which is consistent with recent 
experimental findings \cite{Mathis2017}.
Accordingly to both presented models, the high degree of reversibility
represents an evolutionary advantage in the variable environments 
which disappears when the environments become static and vice versa.

Modeling of environmental changes proposed in this work is 
limited to stylized periodic variations. 
Further insights into the problem 
of variable environments 
can be derived for the stochastic models 
\cite{Charlebois2011,Carja2012,Saeter2015,Hufton2017,Kobayashi2017,Zheng2017,Frey2017}.

The model extension including more general 
asymmetric environmental influence may be laborious, 
but straightforward. It seems that ecology combining   
exogenous and endogenous environmental influence 
(where asymmetry can be a natural side product of the considerations)  
may be more important at the present stage of the modeling.
In our stylized model of cancer cell population, the above
abstract conceptualization is made more instructive 
by defining a periodic environment where exogenous variability
of the effector cell population is considered.

Presented model of heterogeneous tumor growth (Section \ref{sec:Evo})
demonstrates the model-specific therapeutic implications
of evolved bet-hedging strategy and phenotypic plasticity of cancer cells.
The levels of immunotherapeutic efficiency that cause
remarkable differences between genetic and epigenetic entropy measures
become prevalent at late simulation times, where 
mutation changes become very rare.
Calculations show that effective therapies can be expected in the (stable) parametric 
area below a certain threshold of the parameter of therapeutic specificity. 
In difference to, e.g. the model 
\cite{Greulich2012}, 
our model does not make a priori assumption of underlying 
drug resistance. On the other hand, the evolutionary viewpoint based on the concept 
of heterogeneity is a natural source of algebraic
complications which presume more sophisticated 
theoretical treatment.

In general, many dynamic biosystems undergo a fundamental change 
of behavior when one of their parameters passes through 
a threshold value \cite{Sole2016}. The concept of lethal threshold of unstable tumor progression has been 
discussed in \cite{Amor2014,Sardanyes2017}. 
The identification of the threshold as a non-linear 
form of therapeutic effect on fitness is in line with the work \cite{Charlebois2011}.
In \cite{Sewalt2016}, a model of tumor growth is proposed 
where {\em Allee threshold} related to the minimum density 
criterion plays a central role. The ability to manipulate 
selectively the (micro)environment 
with the sustainable intensity and reliable 
detection of the threshold in the therapeutic efficiency 
could be major challenges for therapeutic applications, 
as exemplified may serve immunotherapeutic 
intervention \cite{Li2013}. The concept of environmental manipulation 
has been supported by experimental study \cite{Chen2017} as well
reporting that regular administration of drugs at non-toxic doses
may overcome the drug resistance by shifting 
the therapeutic target from tumor cells to tumor endothelial cells.
This therapeutic strategy is known under the term 
{\em metronomic} chemotherapy and immunotherapy. 
Regarding this, we put into attention 
one of the most findings from  
here presented cancer model, which is a 
positive therapeutic effect achieved 
with less (see also \cite{Lorz2015}) 
when moderate immunotherapeutic specificity (given by $x_p$) is assumed. 

We note, that the original quasispecies model naturally led to the concept of "error threshold" 
for selective competence (see e.g. Fig.\ref{fig:F3} and \cite{Biebricher2005}). 
Consequently, interesting question arises whether the threshold concept 
is transferable to therapeutic context as well.

Regarding the results of the threshold and sensitivity analysis
(see \ref{Ap:Thresh}), the interesting question emerges whether 
carefully chosen linearized model classes which retain
the typical evolutionary properties can be constructed and 
used to analyze the drug resistance observed in 
heterogeneous, evolving and variable cancer cell populations.

\section*{Acknowledgments}
This work was supported by (a)~APVV-15-0485 by Slovak Research and Development Agency; 
(b)~VEGA  No. 1/0250/18; (c) VEGA 1/0156/18 
by Scientific Grant Agency 
of the Ministry 
of Education of Slovak Republic. 

\end{multicols}

\hrule


\begin{figure}[!htb]
\centering
\includegraphics[scale=1.05]{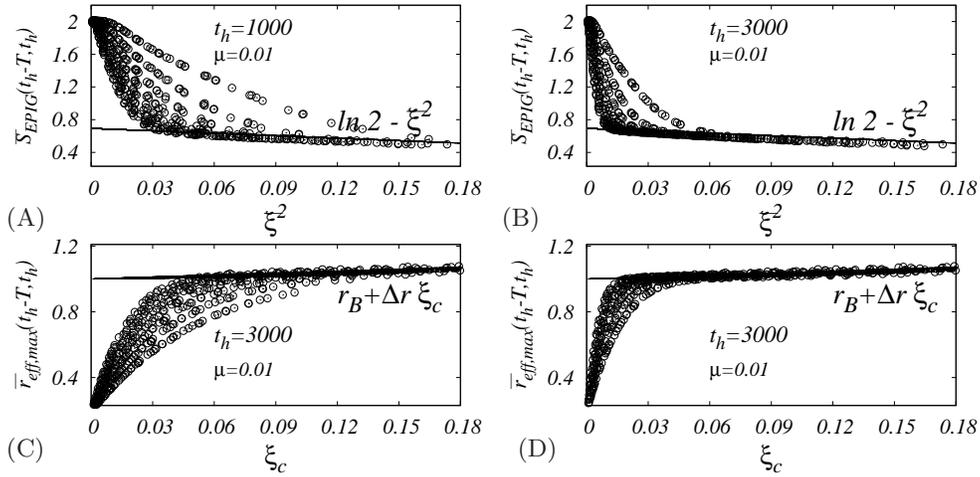} 
\caption{Figure depicts 
results of the sensitivity analysis focused on the 
role of $m$, $T$, $\Delta r$ parameters. 
The results were obtained for the time horizons $t_h\in \{1000, 3000\}$  
[see panels (A),(C) with $t_h=1000$; panels (B),(D) with $t_h=3000$]
and constant $\mu=0.01$.  
Instead of particular dependencies on three parametric 
dimensions represented by $m,T, \Delta r$ (see Fig.\ref{fig:F3}) 
(which lacks holistic perspective), we used 
$\overline{\xi^2}(m,T,\Delta r)$ (see Eq.(\ref{eq:xia2b2})) 
as a single independent variable.
We have studied sensitivity on $\overline{\xi^2}(m,T,\Delta r)$ 
generated by the $(m,T,\Delta r)$ combinations 
with bounds $m \in \langle  0.01,  0.2  \rangle $,  
$T \in \langle 1, 20  \rangle $,  
$\Delta r  \in \langle 0.1, 0.4  \rangle$.
Since $8$ evenly spaced values have been used for each particular variable, 
$8^3=512$ samples were drawn 
from the cubic hyper-lattice. Subsequently, 
each particular $(m,T,\Delta r)$ 
vector was converted to the    
2d projection consisting 
of the coordinates  
$[\overline{\xi^2}(m,T,\Delta r), 
\overline{S}_{EPIG}(t_h-T,t_h)]$, 
where $\overline{S}_{EPIG}(t_h-T,t_h)$ 
$\stackrel{def}{=}$
$(1/T)$ $\sum_{t\in \langle  t_h -T,  t_h \rangle}$
$S_{EPIG}(t)$ [see panels (A), (B)]. 
Here, the summation is performed for the 
period $T$ preceding $t_h$. In analogy with Fig.(\ref{fig:F3}), 
the thresholds reappear.  When the plots (A), (B) 
are compared with the approximate form $\ln 2-\overline{\xi^2}$ (see Eq.(\ref{eq:SEP})), 
particular agreement can be found 
for sufficiently high $\overline{\xi^2}$ (low entropy)  
region where selection process 
for the highest reversibility is more effective. 
The similar tendencies are exhibited by 
the mean replication measure 
$\overline{r\,}_{eff,max}(t_h-T,t_h)$ 
$\stackrel{def}{=}$ $(1/T)$ 
$\sum_{t\in \langle  t_h -T,  t_h \rangle}$ 
$\max_{k=0,  \ldots,  n_s-1}$
$r_{eff}(k,t)$ [see panels (C), (D),  where 
$\xi_c(m,T,\Delta r)$ defined by Eq.(\ref{eq:xiapr}) 
plays a role analogous to that of $\overline{\xi^2}$; 
note that Eq.(\ref{eq:reff}) is used 
to calculate $r_{eff}(k,t)$].  
The long-time asymptotics $r_B+ \Delta r \xi_c$ is given by Eq.(\ref{eq:rBb}). 
In line with the expectations, 
the threshold is more pronounced and the 
relation look sharper at larger 
$t_h$ [panels(B), (D)].}
\label{fig:ApF4}
\end{figure} 

\begin{multicols}{2}

\appendix
\section{Appendices}
\subsection{Long run asymptotics - monoclonal fixation}\label{Ap:Long}

The quasispecies problem does not provide closed-form 
expressions for the solution for a system of many 
species. Nevertheless, numerical results from Subsection \ref{sec:Num} 
stimulated our interest in the asymptotic situation 
$t\gg 1/\mu$ where the limit cycle 
oscillations can be identified. 
We also found that for late-time evolution the 
single clone with the index $n_s/2$ is fixed. 
The case of highly localized (in $k$)
species allows to consider more restrictive version of the constraint 
Eq.(\ref{eq:c0sum}) in the form
\begin{equation}
c(n_s/2,t)=c^{(0)}(n_s/2,t) 
+ c^{(1)}(n_s/2,t) = 1\,. 
\end{equation} 
It is consistent with two 
settings  
\begin{eqnarray} 
c^{(0)}_{\Sigma}(t) 
&\simeq &  
c^{(0)}(n_s/2,t) = 
\frac{1}{2} + \xi(t)\,, 
\label{eq:cxi}
\\
c^{(1)}_{\Sigma}(t)  &\simeq & 
c^{(1)}(n_s/2,t) =
 \frac{1}{2}  - \xi(t)\,, 
\nonumber
\end{eqnarray}  
where the residual plasticity 
is transferred to the 
single auxiliary 
$-1/2<\xi(t) < 1/2 $.
Consequently, using Eq.(\ref{eq:Jswkt})
we obtained $J_{sw}(n_s/2,t) = -\xi$ and the equation for $\xi(t)$ can be 
written as 
\begin{equation}
\frac{d\xi}{dt} =  ( r_0 - \Phi) 
\left(\frac{1}{2}+ \xi\right) - m \xi \,.
\label{eq:dxr0p}
\end{equation} 
Substituting $c^{(0)}(t)$, $c^{(1)}(t)$ from Eq.(\ref{eq:cxi}) 
into Eq.(\ref{eq:reff}) 
we obtained 
\begin{equation}
\Phi(t)  \simeq r_{eff}(t)  = 
r_B +  2  \Delta r \, \xi(t) 
\cos  \left(\frac{2 \pi t}{T}\,\right)\,.
\label{eq:reffcos1}
\end{equation}
Its consequent substitution 
into Eq.~(\ref{eq:dxr0p}) gives
\begin{equation}
\frac{d\xi}{dt} =  
2 \Delta r 
 \left(\frac{1}{4}- \xi^2 \right) \cos\left(\frac{2\pi t}{T}\right)  -  m \xi   \,. 
\label{eq:dxi}
\end{equation}
This nonlinear ODE problem can be solved using the
single harmonic 
approximation (valid for $\xi^2 \ll 1/4$) 
of the limit cycle 
\begin{equation}
\xi(t) \simeq  \xi_{s} \sin\left(
\frac{2\pi t}{T} \right) + 
\xi_{c} 
\cos\left( 
\frac{2\pi t}{T} \right)
\label{eq:xab}
\end{equation} 
with the pair of amplitudes 
\begin{eqnarray} 
\xi_s  
=  
\frac{\pi T\Delta r}{4 \pi^2 + T^2 m^2} \,, 
\qquad 
\xi_c = \frac{m T}{2\pi} \xi_{s} \,\,. 
\label{eq:xiapr}
\end{eqnarray}
The formula clearly 
uncovers 
the relative effects of the processes operating 
at different time 
scales: $1/\Delta r$, 
$1/m$ and $T$. 
The dependence upon $1/\mu$ absents 
due to assumption that clonal selection 
has basically vanished in the long run. 

For the solutions given by Eq.(\ref{eq:xab}), 
the time averaging 
can be simply performed 
for the single period. 
The result is    
\begin{equation}
\overline{\xi^2} \simeq 
\frac{1}{2} \left( \xi_{s}^2 + 
\xi_{c}^2 \right)\,.
\label{eq:xia2b2}
\end{equation}
Similarly, for $T$-periodic $\xi(t)$ 
the mean effective replication rate can be defined by 
\begin{eqnarray}
\overline{r_{eff}} \equiv  \frac{1}{T}  \int_0^T r_{eff} (t) dt\,.
\label{eq:reffT}
\end{eqnarray} 
Then, using Eq.(\ref{eq:reff}), 
Eq.(\ref{eq:cxi}),  Eq.(\ref{eq:reffcos1}), Eq.(\ref{eq:xab})
and Eq.(\ref{eq:reffT}) we obtained
\begin{equation}
\overline{r_{eff}}  =r_B  + \Delta r \, \xi_{c}\,.  
\label{eq:rBb}
\end{equation}
A posteriori confrontation with the condition $\xi^2 \ll 1/4$ 
provides bounding 
$\Delta r  \ll  (4 \pi^2 + T^2 m^2)/ (\pi T \max\{1, mT\})$.
Based on the structure of Eq.(\ref{eq:rBb}), we have proposed 
the numerical analysis 
illustrated by examples in Appendix Fig.\ref{fig:ApF4}(C),(D) 
(see caption of this figure for more details). 
The combination with numerical 
tools in part justifies the 
use of asymptotic approximation presented 
in this Appendix.

We expect potential use of the result described by Eq.(\ref{eq:rBb}) in increasing 
the efficiency of large-scale modeling, performed for the time 
intervals that are much longer than the period of the environment 
contained in the replication rate. A similar type of population-averaged 
growth rate is discussed in \cite{Thattai2004} within the context 
of gene expression of bacterial populations exposed to environmental variations.

Assuming that the only clone (the winner) survives,  we get a trivial limit 
${S_G|}_{t\rightarrow \infty} \rightarrow 0$. 
However, the epigenetic alterations have non-trivial consequences.
If Eq.(\ref{eq:cxi}) is substituted 
into Eq.(\ref{eq:SEPIG0}), one obtains
{\small
\begin{eqnarray}
S_{EPIG}(t)\simeq -  
\sum_{j\in \{-1, 1\}} \bigg(\frac{1}{2} + j \xi(t)\bigg)
\ln 
\bigg(\frac{1}{2} 
+ j \xi(t)\bigg)\,.  
\label{eq:SEP}
\end{eqnarray}}
The formula can be analyzed by calculating its time averages. For this aim, 
the Taylor series of the order 
${\mathcal O}(\xi^2)$ can be used. The entropy averaged over the period 
$\overline{S}_{EPIG} $ $\simeq $ $ \ln 2 $ $- \overline{\xi^2}$ 
used in combination with Eq.(\ref{eq:xiapr}) and Eq.(\ref{eq:xia2b2})
provides $(\overline{S}_{EPIG} - \ln 2)  \sim (\,- \Delta r^2)$ 
in a qualitative agreement 
with the simulation trend shown in Fig.\ref{fig:F2}(A),(D),(E),(F), 
Fig.\ref{fig:F3}(B) and also short 
sensitivity study reported in 
Appendix Fig.\ref{fig:ApF4}(A),(B).  

It is clear that the calculations of the characteristics 
for separate variables are not universal across parametric values. 
The first step towards integrated view is manifested 
in Fig.\ref{fig:ApF4}, where the time averaged $S_{EPIG}(t)$ 
and $\max_k r_{eff}(k,t)$ are plotted versus  
auxiliary variables $\overline{\xi^2}$ and $\xi_c$ 
[see Eq.(\ref{eq:xiapr}) 
and Eq.(\ref{eq:xia2b2}) below] 
for different combinations of 
$m, T, \Delta r$ inputs. The explicit forms of  
$\overline{\xi^2}(m,T,\Delta r)$ 
and $\xi_c(m,T,\Delta r)$ follow 
from the asymptotic results  
for the clone selection problem represented by Eq.\ref{eq:rBb}.

Contrary to previous considerations let us abandon for a moment 
predetermination of $\varphi(k_{\rm central})$. Instead, 
we discuss an alternative formulation that assumes equivalence of the asymptotic 
evolutionary solution with the one-dimensional 
static maximum of the mean 
$\overline{r_{eff}}(\varphi)$.
Such analysis is suitable 
for late times, when selection
process is completed and, 
consequently, $k$-dependency of 
$\varphi(k)$ becomes irrelevant. In that case, 
the parametrization assuming the real-valued 
$\varphi \in \langle 0, 1\rangle$ can be used
and subsequent optimization performed.

In analogy with Eq.(\ref{eq:cxi}) we assume  
$c^{(z)}(\varphi,t)=1/2 + (-1)^z \tilde{\xi}(\varphi,t)$.
Continuous $\tilde{\xi}(\varphi,t)$ 
dependence may serve for the selection of the 
$\varphi_{opt}\in \langle 0,1 \rangle$ 
which solves the optimization task 
$\varphi_{opt}=\arg\max_{\varphi\in \langle 0,1\rangle} 
\overline{\widetilde{r_{eff}}}(\varphi)$. 
When $\varphi$ remains unspecified, 
the original Eq.(\ref{eq:dxi}) may be rewritten to the form
\begin{eqnarray}
\frac{d\tilde{\xi}}{dt}= 
\frac{\Delta r}{2} (1-4 \tilde{\xi}^2)
\cos\left(\frac{2\pi t}{T}\right) 
+  m \left(\frac{1}{2} - 
\tilde{\xi}-\varphi\,\right)\,. 
\label{eq:tilxi}
\end{eqnarray} 
In analogy with Eq.(\ref{eq:reffcos1}) 
we define $\widetilde{r_{eff}}(\varphi,t)= 
r_B+ 2 \Delta r \tilde{\xi} 
\cos \left(2\pi t/T\right)$ 
and the corresponding time average 
$\overline{\widetilde{r_{eff}}}(\varphi)$.
 Obviously, the conflict between discrete $\varphi(k_{\rm central})$ 
and the solution of the continuous problem, 
$\varphi_{opt}$,
can occur if the list of the possible discrete 
values $\{\varphi(k)\}_{k=0}^{n_s-1}$ 
does not contain $\varphi_{opt}$.

Nowadays, stochastic model variants become relatively independent 
field 
of research. For the lack of space, below we outline only 
an elementary model based on stochastic modification of the periodic environmental model. 
The reasons are as it follows: 
(i)~recent studies point to 
the extrinsic noise as a factor influencing 
genetic and epigenetic determinants of the replication; 
(ii)~there is an increased interest in a description 
of the intermediate environments between strictly 
periodic and noise processes;  
(iii)~the noise responses can be considered 
as a suitable stability indicators.

Our elementary stochastic extension is based on the generalization 
in which the original regular $\cos(2 \pi t/T)$ function 
from Eq.(\ref{eq:tilxi}) is replaced by the periodized 
variant $w_{ou}(t)$ of the continuous Ornstein-Uhlenbeck 
random continuous process described by  
\begin{eqnarray} 
d w_{ou} =  \frac{1}{\tau_{ou}}
\left( \cos\left(\frac{2\pi t}{T}\right) 
-  w_{ou}\,\right)  +  \sigma_{ou} d W\,,    
\end{eqnarray}
where the constant $\sigma_{ou}$ 
changes the original unit dispersion of the standard 
Wiener process $W(t)$. The basic idea of the 
process is that relaxation to 
the instantaneous $\cos(2 \pi t/T)$ is described 
by the relaxation time $\tau_{ou}$. 
Moreover, as the periodization retains the influence of the former  
time scale $T$, different 
levels of stochasticity can be simulated
using $\tau_{ou}$. 

In accordance 
with the structure of Eq.(\ref{eq:tilxi}), 
new $w_{ou}(t)$ term is used to modify 
\begin{eqnarray}
\frac{d \xi_{ou}}{dt} 
&=& 
\frac{\Delta r}{2}\,\left(\, 1-4 \xi_{ou}^2\right)\, 
w_{ou}(t) 
\label{eq:xiouwou}
\\
&+&  m\left( \frac{1}{2} - \xi_{ou} - \varphi\,\right) 
\nonumber
\end{eqnarray}
as well as the effective replication rate  
$r_{ou-eff} =  r_B + 2  \Delta r \xi_{ou} w_{ou}$. 
Again, to determine  the optimum uniquely, 
the quantity $\overline{r_{ou-eff}}(\varphi)$ 
needs to be investigated for the late times 
where the effects of initial conditions 
become negligible.

\begin{figure}[!htb]
\centering
\includegraphics[scale=1.02]{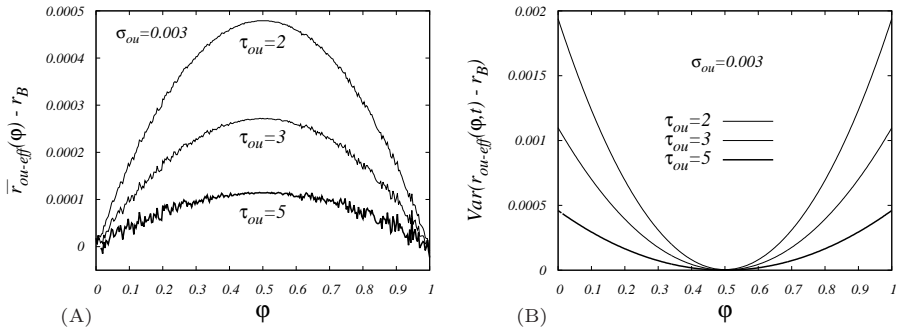} 
\caption{The effective 
replication rate and corresponding fluctuations calculated 
for the parameters $T=10$, $\Delta r=0.1$, $m=0.1$ 
using Eq.(\ref{eq:xiouwou}). 
To attain stationary regime, 
initial 200 periods were discarded.  
Next thousand periods were used to calculate the mean values.
Panel A confirms the optimality 
of the mean clonal replication 
for $\varphi=1/2$ is confirmed. 
According to panel (B), 
the largest fluctuations are observed 
for the boundary region.} 
\label{fig:ApF4X}
\end{figure}

As depicted in Fig.\ref{fig:ApF4X}), 
the solution of the initial value problem obtained 
for given $\varphi$ can be used to calculate 
$\overline{\widetilde{r_{ou-eff}}}(\varphi)$. 
Since we are focussed on the region of $\tau_{ou}\ll T$ and 
very small $\sigma_{ou}$, there is 
no need for separate presentation related 
to deterministic limit.  Two claims could be made regarding 
$w_{ou}$, $\xi_{ou}$ 
and $r_{ou-eff}$ properties: 
(a)~Because the relaxing of $w_{ou}(t)$ is unable to reach 
the heights of harmonic amplitudes, the process 
$\xi_{ou}(t)$ causes that $\overline{r_{ou-eff}}(\varphi)$ 
values are systematically smaller than their deterministic 
counterparts; 
(b)~The fluctuations characterized by the variance 
$\mbox{Var}(r_{ou-eff}(\varphi,t)-r_B)$ 
show an extremal 
impact of the highly unstable boundary regions 
$\varphi \rightarrow 0^{+}$ 
and $\varphi \rightarrow 1^{-}$.  
\end{multicols}

\begin{figure}
\centering 
\includegraphics[scale=1.00]{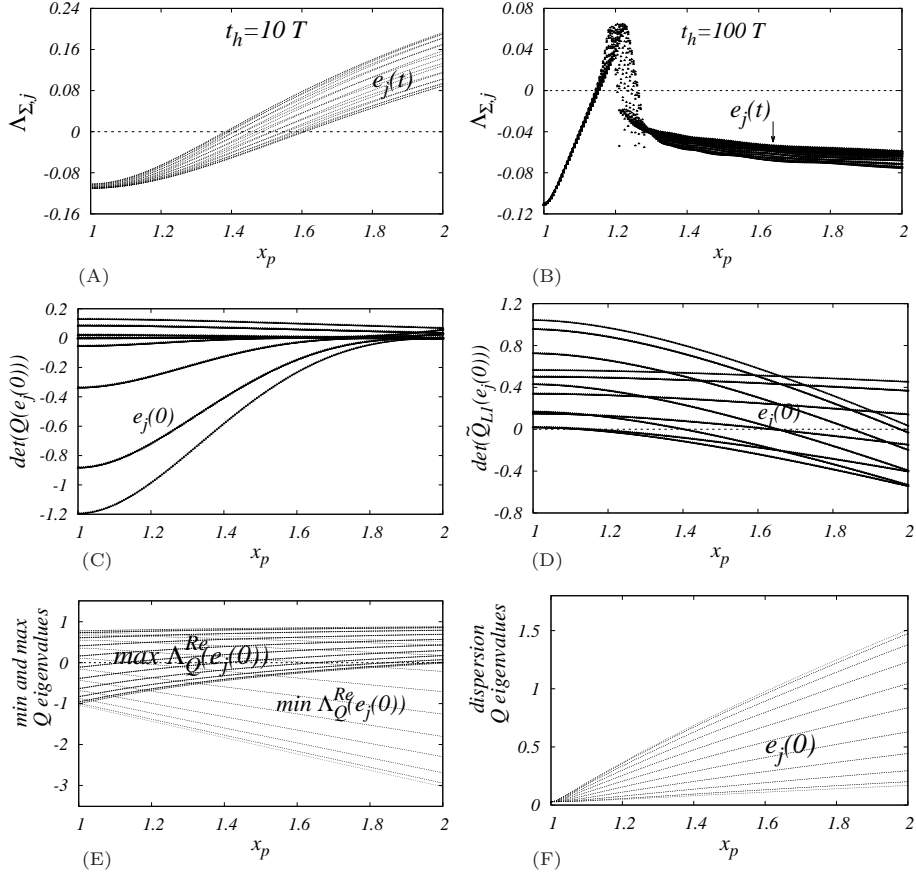}
\caption{Auxiliary calculations that allow us to better understand 
the complexities around the threshold of the parameter 
$x_p$. Calculations, as in many previous cases, were performed 
for parameters $a=1$, $r=1$, $m=0.05$, $\mu=0.01$, 
$T=10$, $g=0.45$,  
$e_B=0.5$, $\Delta e=0.4$, 
$\varphi_L=0.25$, $C_{P0}=0.001$. 
The new parameter $\delta C_{P0}=0.0001$ [see Eq.(\ref{eq:slightdifini})] 
characterizes initial separation of trajectories; 
$N_e=20$ variants of the phase shifted $e_j(t)$ and $e_j(0)$ have been taken 
into account to characterize the sensitivity. 
Panels (A), (B) show how the growth instability is manifested by the Lyapunov-like exponents $\Lambda_{\Sigma,j}$,  $j=1, 2,\ldots, N_e$. 
We see that the influence of phase shifts is getting 
weaker as the time is passing [shadowing of initial condition Eq.(\ref{eq:InitC}) where the sole nonzero 
abundance is $C_P(0)$]. For the purposes of alternative approach, we 
consider panels (C) and (D) including
$det(Q(e_j(0)))$ (as a product of eigenvalues)
and $det({\tilde{Q}}_{L1}(e_j(0)))$, respectively.
Interestingly, there are many mutual crossings in (D) starting above $x_p\sim 1.2$.  
Panel (E) shows maximal and minimal elements of the set 
${\bm\Lambda}_{Q}^{Re}(e_j(0))$ (determined 
for each $e_j(0)$ separately).  Alternatively, the system of 
dispersions in panel (F) helps us to 
characterize the distribution of eigenvalues arranged in
${\bm\Lambda}_{Q}^{Re}(e_j(0))$.} 
\label{fig:F7}
\end{figure}

\hrule

\begin{multicols}{2}

\subsection{Therapeutic thre\-shold, 
ini\-tial stages of tumor formation, 
sen\-sitivity analysis}\label{Ap:Thresh}

To conceive better the effect of the therapeutic threshold
as well as to support generality of the suggested concept (for $ x_p $ parameter), 
we analyze initial stages of the tumor 
formation in more detail. 
Additionally to our previous layout,
below we focus on the sensitivity analysis in the space
of therapeutic options or available uncertain inputs.
For that aim, we formulate the threshold problem 
in a way enabling us to carry out
suitable 
analytical or semi-analytical approximations.

Although the issues of uncertainty and risk are
not typically addressed within the ODE, the relevant information 
in this context can be obtained by studying assembly of alternative 
phase shifts that are exogenous inputs of our ODE system.
Assuming that medical treatment will be 
confronted with an uncertainty in the abundance of the effector
cell population ($\sim \Delta e$), 
the additional phase variability has been
incorporated by modifying the original periodic model of $e(t)$. 
In the modified version, for 
every simulation run 
$j=1,2,\ldots, N_e$
of the duration $t_h$ the specific 
$j$-dependent phase shift $2 \pi j /N_e$ is introduced as an extra degree of freedom with the corresponding abundance    
\begin{equation}
e_j(t)=e_B  + \Delta e \cos\bigg[ \,  2 \pi \bigg(  
\frac{j}{N_e} +  \frac{t}{T}\,\bigg)\, \bigg]
\end{equation} 
on the interval $t \in \langle 0, t_h  \rangle$.
In such case, dynamics of $e_j(t)$ influences the respective 
population dynamics through 
$\propto -\tilde{\alpha} a e_j(t) \tilde{C}/(C_{\Sigma,j}+g)$ term 
modifying the original form of Eq.(\ref{eq:Rform}). 
The sensitivity to the phase differences provide 
additional information related to the system 
trajectories. 

We assume that within the context 
of here studied growth phenomena,
the relevant information can be described by the 
Lyapunov-like exponent
\begin{equation} 
\Lambda_{\Sigma,j}= \frac{1}{t_h}
\ln \frac{ \left| C_{\Sigma,j}^{[1]}(t_h) -
C_{\Sigma,j}^{[2]}(t_h)
\right| }{|{\delta C}_{P0}|}\,.
\end{equation}
defined by means of two 
horizon values $C_{\Sigma,j}^{[1]}(t_h)$, 
$C_{\Sigma,j}^{[2]}(t_h)$ 
[splitting the original $C_{\Sigma}$ 
from Eq.(\ref{eq:CSig})]
obtained under the action of $e_j(t)$ for two initial conditions 
\begin{eqnarray}
&& {\rm \mbox{\small path}}
\,\,\, C^{[1]}_{\Sigma,j}(t):  \,\,\,
{\rm \mbox{\small initial}}\,\,  
C_P(0)=C_{P0}\,,\hfill
\label{eq:slightdifini}
\\
&& {\rm \mbox{\small path}} \,\, C^{[2]}_{\Sigma,j}(t): \,\,\,
{\rm \mbox{\small initial}}\,\,\, 
C_P(0)=C_{P0}+\delta C_{P0}\,,\hfill
\nonumber
\end{eqnarray}
where $\delta C_{P0}$ is small shift in $C_{P0}$ defined by Eq.(\ref{eq:InitC}).  

Since the system of phenotypes 
begins to evolve from a small monoclonal primary tumor, 
there is no initial heterogeneity (see Fig.\ref{fig:F2} and Fig.\ref{fig:F6}).
Thus, for small times the dynamics of species 
abundances allows for 
a linear analysis that offers some computational benefits. 
This section continues with the development 
of semi-analytical approximation which supports relevance of the threshold approach.  
The aim is to extend this intuitive concept
by examining stability of the initial conditions quantified
by the spectrum of the  local Lyapunov exponents.

Within the applied quasi-static approximation each $e_j(t)$, $j=1,2,\ldots, N_e$ 
is replaced by the constant value $e_j(0)=e_B+\Delta e \cos(2 \pi j/N_e)$. In such 
elementary case, Eq.(\ref{eq:dCPdt}) can be linearized 
in the vicinity of the initial condition given by Eq.(\ref{eq:InitC}). 
The linearization provides an approximate solution with the dynamic "fastest" mode 
written in the matrix form $ \exp({\bf Q}(e_j(0)) t) $, 
where ${\bf Q}(e_j(0))$ denotes the local Jacobian matrix of the system
Eq.(\ref{eq:dCPdt}) we specify in the following. Then,
the diagonalization of ${\bf Q}(e_j(0))$ can be performed and 
the stability of the initial condition 
can be checked by investigating real parts of eigenvalues  
\begin{equation}  
{\bm\Lambda}^{Re}_{Q,j}\equiv \Re[ \mbox{diag}({\bf Q}(e_{j}(0)) ]\, 
\end{equation}
Then, by studying $N_e$ variants of ${\bm\Lambda}^{Re}_{Q,j}$ spectra   
the possible impacts of the therapy on the very early stages 
of tumor formation can be obtained.  

For later purposes, we introduce auxiliary variables 
\begin{eqnarray}
r_{P0} &=& r (1-C_{P0})\,,  
\,\,\,
a_{e} = \frac{a e_j(0) g}{(C_{P0}+g)^2}\,, 
\\
Q_P &=& r_{P0} - r C_{P0} -  a_{e} -  6 \mu\,,  
\nonumber 
\\
Q_{ae} &=& ( a_{e}  -  r ) C_{P0}\,.  
\nonumber 
\end{eqnarray} 
The resulting ${\bf Q}$ can be ordered into the blocks 
{\small
\begin{eqnarray}
\mbox{\large ${\bf Q}$}
=
\left[ \begin{array}{c:c:c:c}
Q_P        &   (Q_{ae} ,\, Q_{ae})        & (Q_{ae} ,\, Q_{ae})          
&   (Q_{ae} ,\, Q_{ae})   \\
\hdashline            
&&&    \\  
\binom{\mu}{\mu}   &  \tilde{\bf Q}_{L0} &  {\bf 0}_{2\times 2}
           &   
{\bf 0}_{2\times 2}    \\
\hdashline   
&&&    \\   
\binom{\mu}{\mu} &  {\bf 0}_{2\times 2} &  \tilde{\bf Q}_{L1}   
&  
{\bf 0}_{2\times 2}   \\  
\hdashline  
&&&    \\  
\binom{\mu}{\mu}  &  {\bf 0}_{2\times 2}  &  {\bf 0}_{2\times 2}  
&  
\tilde{\bf Q}_H   \\ 
\hdashline
\end{array}\right]\,.
\end{eqnarray}} 

Here, the evolutionary mechanism is highlighted 
by the leftmost column which mediates 
the interconnection of species. 
The first row including non-diagonal 
$Q_{ae}$ elements simply reflects the immune feedback.
Three $2 \times 2$ sub-matrices of the type  
\begin{eqnarray}
\tilde{\bf Q}_{X} 
=  
\left(
\begin{array}{lc}
Q^{(0)}_{X}  &  
m (1-\varphi_{L})   
\\ 
m \varphi_L   &  Q^{(1)}_{X}     
\end{array} \right)\,, 
\,\, X \in \{L0,L1,H\} \
\end{eqnarray}
correspond to the respective clones $L0, L1, H$.
Their switching capability is evident from the 
nondiagonal structure of $\tilde{\bf Q}_{X}$.
The interplay of the proliferative, switching, 
immune and therapeutic effects is 
compactly expressed in the following diagonal terms 
$Q^{(z)}_{X}$, $z\in\{0,1\}$. For the phenotypic alternative $z = 0$ 
we have 
{\small
\begin{eqnarray}
\left(  
\begin{array}{c} 
Q^{(0)}_{L0}   \\  \vspace*{-2.9mm} \\
Q^{(0)}_{L1}   \\   \vspace*{-2.9mm} \\
Q^{(0)}_{H} 
\end{array} \right) = 
r_{P0}  - m  \left( 
\begin{array}{c} 
\varphi_L    \\  \vspace*{-2.9mm} \\
\varphi_L    \\  \vspace*{-2.9mm} \\
\frac{1}{2}   \\  
\end{array}
\right)  -  \frac{a e_j(0)}{C_{P0}+g} \left( 
\begin{array}{c} 
x_p                  \\ \vspace*{-2.9mm} \\
\frac{1}{x_p}    \\ \vspace*{-2.9mm} \\ 
x_p   
\end{array}
\right)\, , 
\label{eq:Q0LH}  
\end{eqnarray}}
whereas in the $z = 1$ case we obtained 
{\small
\begin{eqnarray}
\left(  
\begin{array}{c} 
Q^{(1)}_{L0}   \\  \vspace*{-2.9mm} \\
Q^{(1)}_{L1}   \\   \vspace*{-2.9mm} \\
Q^{(1)}_{H} 
\end{array} \right) 
=  r_{P0}  - m  \left( 
\begin{array}{c} 
1-\varphi_L    \\   \vspace*{-2.9mm} \\
1-\varphi_L    \\   \vspace*{-2.9mm} \\
\frac{1}{2}      \\  
\end{array}
\right)  -  \frac{a e_j(0)}{C_{P0}+g} 
\left( 
\begin{array}{c} 
\frac{1}{x_p}                  \\ \vspace*{-2.9mm} \\
x_p       \\ \vspace*{-2.9mm} \\ 
\frac{1}{x_p}   
\end{array} \right)\,.
\label{eq:Q1LH}   
\end{eqnarray}}
The methodology offers interpretation in which the threshold of $x_p$
separates the phase of small and large instability. 
The results of numerical calculations are shown in Fig.\ref{fig:F7}. 
Important general finding is, that the threshold behavior can be manifested 
not only by the nonlocal measures (e.g. using $\Lambda_{\Sigma,j}$), 
but by the calculations based on ${\bf Q}(e_j(0))$ as well.
The long-term ($t_h=100 T$) dependence $\Lambda_{\Sigma,j}(x_p)$ 
depicted in Fig.\ref{fig:F7}(B) seems to be fairly consistent 
with roughly estimated threshold $x_{p,{\rm thr}} \in (1.12, \,1.14)$ 
identifiable from Fig.\ref{fig:F5}(H). In addition, regarding sensitivity to $e_j(t)$, 
Fig.\ref{fig:F7}(B) shows that after a long period, the consequences
of the phase differences gradually vanish due to convergence
onto the universal (in the sense of its phase-independence) limit
cycle attractor. Conversely, the panel (A), where $ t_h=10 T$,
shows no more evidence for the threshold-type behavior 
in the terms $\Lambda_{\Sigma,j}$. Regarding evaluation
of the set of the measures derived from ${\bf Q}(e_j(0))$,
it seems that detailed enough knowledge available from the local
characteristics derived from ${\bf Q}(e_j(0))$ makes it possible
to substitute the lack of information.

The local characteristics based on ${\bf Q}(e_j(0))$
show capacity to unmask
expanding or contracting tendencies that vary 
with $x_p$. While $det({\bf Q}(e_j(0)))$ 
depicted in the panels (C), (D) indicates that system of trajectories 
may begin to expand due to $x_p$, the minors 
$det({\bf Q}_{L1}(e_j(0)))$,  
$j=1,2, \ldots, N_e$,
on the contrary, indicate a shift towards more contractive effects
(we verified that minors $det(\tilde{\bf Q}_{L0})$ and 
$det(\tilde{\bf Q}_{H})$ confirm the same tendency).
It means, that therapeutically induced 
separation of eigenvalues   
$\bm{\Lambda}^{Re}_{Q,j}$ 
[see panels (E), (F) of Fig.\ref{fig:F7}\,] 
cannot be simply 
attributed to diversity of subpopulations, but, more likely, it reflects
diversity of the replication rates 
causing the rapid clonal expansion 
with possible dominance of fittest species. 	
\end{multicols}

\clearpage

\begin{multicols}{2}

\end{multicols}
\end{document}